\documentclass[english,prl,twocolumn, aps,amssymb,footinbib,showpacs]{revtex4-1}
\usepackage[T1]{fontenc}
\usepackage{amsmath}
\usepackage{amssymb}
\usepackage{graphicx}
\usepackage{braket}
\usepackage{babel}
\usepackage{color}
\usepackage[colorlinks]{hyperref}

\makeatletter

\@ifundefined{textcolor}{}
{
 \definecolor{BLACK}{gray}{0}
 \definecolor{WHITE}{gray}{1}
 \definecolor{RED}{rgb}{1,0,0}
 \definecolor{GREEN}{rgb}{0,1,0}
 \definecolor{BLUE}{rgb}{0,0,1}
 \definecolor{CYAN}{cmyk}{1,0,0,0}
 \definecolor{MAGENTA}{cmyk}{0,1,0,0}
 \definecolor{YELLOW}{cmyk}{0,0,1,0}
 }

\usepackage{bm}\@ifundefined{definecolor}{\usepackage{color}}{}

\newcommand{\re}[1]{\text{Re}\!\left(#1\right)}
\newcommand{\im}[1]{\text{Im}\!\left(#1\right)}

\newcommand{\showfontsize}{\f@size{} pt}
\newcommand{\kzero}[0]{\ket{0}_\alpha}
\newcommand{\kone}[0]{\ket{1}_\alpha}
\newcommand{\ba}[0]{\bm{a}}
\newcommand{\bb}[0]{\bm{b}}
\newcommand{\bq}[0]{\bm{q}}
\newcommand{\bvarphi}[0]{\bm{\varphi}}
\newcommand{\kappac}[0]{\kappa_\text{conf}}
\newcommand{\kappae}[0]{\kappa_\text{err}}
\newcommand{\kun}[0]{\kappa_a}
\newcommand{\kde}[0]{\kappa_2}
\newcommand{\ainf}[0]{\alpha_{\infty}}

\usepackage[T1]{fontenc}
\usepackage[latin9]{inputenc}
\usepackage{textcomp}
\usepackage{color}

\newcommand{\beginsupplement}{%
        \setcounter{table}{0}
        \renewcommand{\thetable}{S\arabic{table}}%
        \setcounter{figure}{0}
        \renewcommand{\thefigure}{S\arabic{figure}}%
        \setcounter{equation}{0}
        \renewcommand{\theequation}{S\arabic{equation}}
     }

\usepackage{babel}

\makeatother

\begin{document}

\title{Exponential suppression of bit-flips in a qubit encoded in an oscillator}

\author{Rapha\"el Lescanne$^{1,2}$, Marius Villiers$^{1,2}$, Th\'eau Peronnin$^{3}$, Alain Sarlette$^{2}$, Matthieu Delbecq$^{1}$, Benjamin Huard$^{3}$, Takis Kontos$^{1}$, Mazyar Mirrahimi$^{2}$, Zaki Leghtas$^{4, 1, 2}$}
\affiliation{$^1$Laboratoire de Physique de l'Ecole Normale Sup\'erieure, ENS, Universit\'e PSL, CNRS, Sorbonne Universit\'e, Universit\'e Paris-Diderot, Sorbonne Paris Cit\'e, Paris, France}
\affiliation{$^2$QUANTIC team, INRIA de Paris, 2 Rue Simone Iff, 75012 Paris, France}
\affiliation{$^3$Universit\'e Lyon, ENS de Lyon, Universit\'ee Claude Bernard Lyon 1, CNRS, Laboratoire de Physique, F-69342 Lyon, France}
\affiliation{$^4$Centre Automatique et Syst\`emes, Mines-ParisTech, PSL Research University, 60, bd Saint-Michel, 75006 Paris, France}

\begin{abstract}
{A quantum system interacts with its environment, if ever so slightly, no matter how much care is put into isolating it. As a consequence, quantum bits (qubits) undergo errors, putting dauntingly difficult constraints on the hardware suitable for quantum computation. New strategies are emerging to circumvent this problem by encoding a qubit non-locally across the phase space of a physical system. Since most sources of decoherence are due to local fluctuations, the foundational promise is to exponentially suppress errors by increasing a measure of this non-locality. Prominent examples are topological qubits which delocalize quantum information over real space and where spatial extent measures non-locality. In this work, we encode a qubit in the field quadrature space of a superconducting resonator endowed with a special mechanism that dissipates photons in pairs. This process pins down two computational states to separate locations in phase space. As we increase this separation, we measure an exponential decrease of the bit-flip rate while only linearly increasing the phase-flip rate. Since bit-flips are continuously and autonomously corrected at the single qubit level, only phase-flips are left to be corrected via a one-dimensional quantum error correction code. This exponential scaling demonstrates that resonators with non-linear dissipation are promising building blocks for universal fault-tolerant quantum computation with drastically reduced hardware overhead.}

\end{abstract}
\date{\today}
\maketitle

Protecting quantum states against decoherence is a fundamental problem in physics, and is pivotal for the future of quantum computing. The theory of quantum error correction (QEC) and its fault-tolerant implementation \cite{Shor1995, Steane1996} provides a solution. In QEC, groups of noisy physical qubits are arranged together to encode qubits with reduced noise, and fault-tolerance establishes that noisy quantum computers can operate reliably if the noise is below a threshold. A strong focus in quantum architecture design has been to increase this threshold to a value within experimental reach, {but the required hardware overhead remains daunting} \cite{Fowler2012}. Therefore, there is a pressing need for new ideas to encode and protect quantum information.

Let us start by understanding why classical information is so stable. Consider a light switch, which has two stable states labeled 0 and 1. Their stability is provided by two properties. First, in order to change states one needs to apply a force to overcome an energy barrier, usually provided by the deformation of a spring. Second, friction between mechanical parts is essential for stability: when a perturbation randomly deviates the switch from its stable state, the gained entropy must be dissipated into a reservoir in order to {recover} the initial state. Can these two properties be transposed to protect quantum information?

{The $\ket{0}$ and $\ket{1}$ states of a qubit}, such as electronic orbitals of an ion or energy levels of a non-linear resonator, {often} have overlapping supports in phase space. First, one needs to isolate the two states so that they no longer overlap \cite{Fluhmann2019, Campagne2019} and separate them by an energy barrier \cite{Brooks2013, Albrecht2016, Lin2018, Earnest2018, Smith2019, Puri2017}. The second property, friction (or dissipation)  leaks information about the system and therefore seems incompatible with the requirement for a qubit to adopt quantum superpositions of states. 
Remarkably, there exists a dissipative mechanism, known as two-photon dissipation, which stabilizes the $\ket{0}$ and $\ket{1}$ states of a qubit without affecting quantum superpositions of the two \cite{Wolinsky1988}. 

{Recent superconducting circuit experiments \cite{Leghtas2015, Touzard2018a} have demonstrated that a resonator endowed with two-photon dissipation develops a manifold of steady states spanned by two states $\kzero$ and $\kone$, lying in two distinct locations of the resonator two-dimensional (2D) phase space. The combination of dissipation and non-locality should prevent random swaps between $\kzero$ and $\kone$ (bit-flips). However, the circuit architectures mediating the two-photon dissipation impinged errors on the resonator. These experiments fell short of crossing the demanding threshold where the correction is faster than the occurrence of all errors, including those induced by the correcting mechanism itself.}

\begin{figure}
\includegraphics[width=\columnwidth]{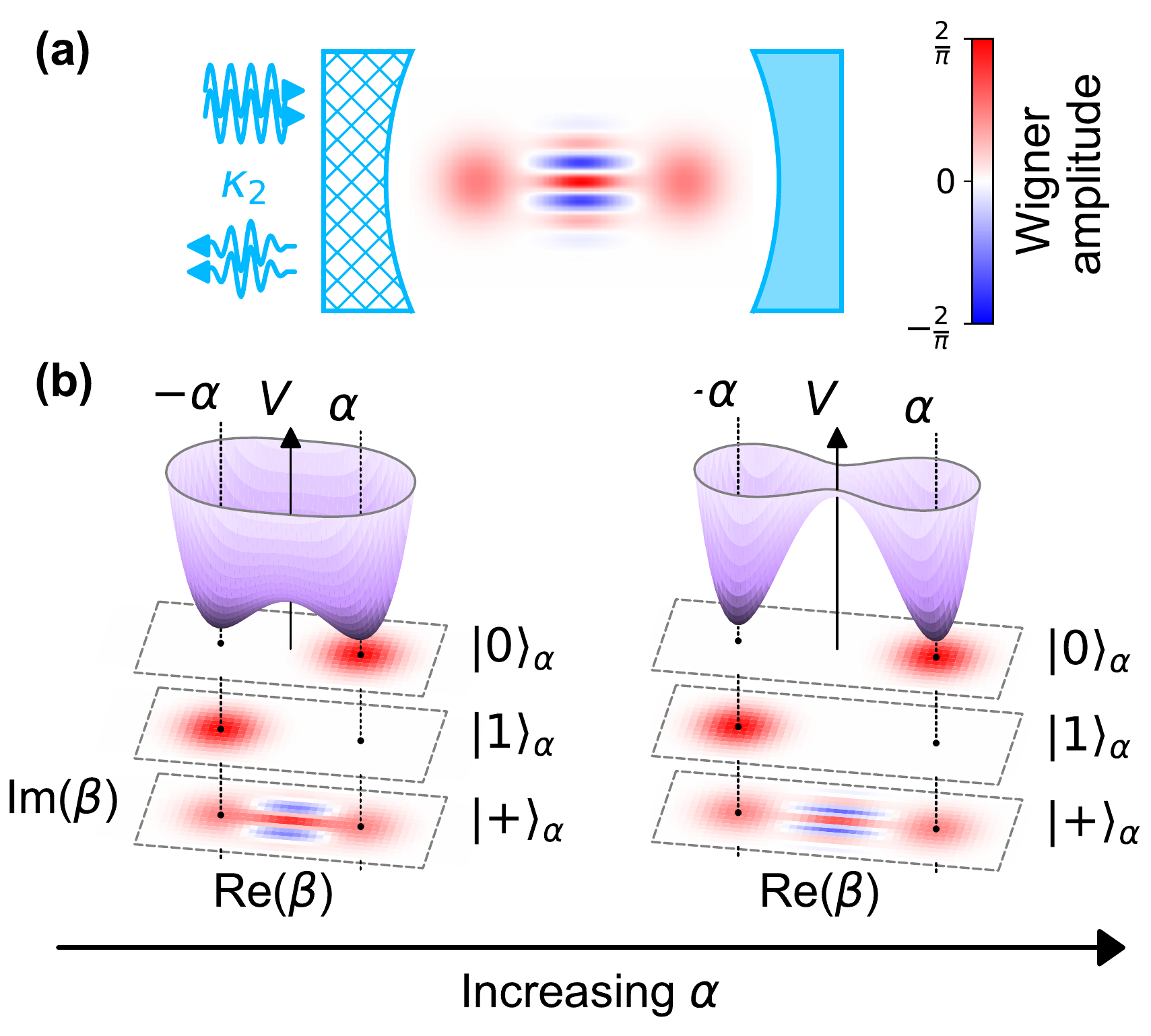}
\caption{\textbf{The cat-qubit} (\textbf{a}) Quantum information is encoded in a resonator (blue mirrors) coupled to its environment through a special apparatus (hatched mirror) where pairs of photons are exchanged at rate $\kappa_2$ (double arrows). (\textbf{b}) This dynamics is illustrated by a pseudo-potential $V$ (purple) defined over the resonator {quadrature} phase space ($\beta$ plane). The cat-qubit states $\kzero$ and $\kone$ lie in the minima of $V$ and are separated in phase space as shown by their Wigner representations (stacked color plots). Bit-flip errors, which randomly swap $\kzero$ and $\kone$, are exponentially suppressed by increasing this separation. {Crucially, } this pseudo-potential does not alter quantum superpositions {of $\kzero$ and $\kone$} such as the Schr\"odinger cat state $\ket{+}_\alpha$.}
\label{fig1}
\end{figure}

In this work, we measure an exponential {decrease of the bit-flip rate} as we increase the separation between states {$\kzero$ and $\kone$}, while only linearly increasing the phase-flip rate (errors {that} scramble the phase of a superposition of $\kzero$ and $\kone$). The bit-flip time reaches 1~ms, a 300-fold improvement over the energy decay time of the resonator. This was made possible by inventing a circuit which mediates a pristine non-linear coupling between the resonator and its environment, {thus circumventing the problems of previous implementations \cite{Leghtas2015,Touzard2018a}}. Our qubit combines two unique features: only phase-flips remain to be actively corrected \cite{Guillaud2019}, and its 2D phase space can be accessed to perform gates \cite{Mirrahimi2014, Grimm2019, Puri2019, Guillaud2019}, making it an ideal building block for scalable fault-tolerant quantum computation with a significant reduction in hardware overhead \cite{Guillaud2019}. 


We follow the paradigm of cat-qubits \cite{Leghtas2013, Mirrahimi2014} where information is encoded in quantum superpositions of resonator states (see Fig.~\ref{fig1}):
\begin{eqnarray*}
\ket{0}_\alpha &=& \frac{1}{\sqrt{2}}\left(\ket{+}_\alpha+\ket{-}_\alpha\right) = \ket{+\alpha} + \mathcal{O}(e^{-2|\alpha|^2})\\
\ket{1}_\alpha &=& \frac{1}{\sqrt{2}}\left(\ket{+}_\alpha-\ket{-}_\alpha\right) = \ket{-\alpha} + \mathcal{O}(e^{-2|\alpha|^2})
\label{eq:01}
\end{eqnarray*}
where $\ket{\pm}_\alpha=\mathcal{N_\pm}\left(\ket{\alpha}\pm\ket{-\alpha}\right)$, $\ket{\alpha}$ is a coherent state with complex amplitude $\alpha$, and $\mathcal{N_\pm}=1/\sqrt{2(1\pm e^{-2|\alpha|^2})}$. All these states contain an average number of photons $\approx|\alpha|^2$ for $|\alpha| > 1$.
A significant source of errors in a resonator is energy decay which collapses all states ($\kzero$ and $\kone$ included) towards the vacuum, thus erasing any encoded information. This decay is balanced by a mechanism where the resonator exchanges only pairs of photons with its environment (Fig.~\ref{fig1}a) \cite{Wolinsky1988}, known as two photon dissipation.
This dynamics is modeled by the following loss operator
\begin{equation}
\label{eq:H2}
    \bm{L}_2 = \sqrt{\kappa_2}\left(\ba^2-\alpha^2\right)\,,
\end{equation}
where $\ba$ is the annihilation operator of the resonator, $\kappa_2$ {is the rate at which pairs of photons are exchanged with the environment} and the term in $\alpha^2$ results from a drive which inserts pairs of photons \cite{supplement}. The cat-qubit states $\kzero$, $\kone$ and all their superpositions are steady states of this dynamics. A convenient tool to visualize the {semi-classical dynamics of} \eqref{eq:H2} is the pseudo-potential $V$ defined over the complex plane as
$-\nabla V(\beta)=\frac{d\beta}{dt}$, 
where $\beta$ is the expectation value of $\ba$ at time $t$ in a semi-classical approximation \cite{supplement}. Stable steady states are local minima of $V$ (see Fig.~\ref{fig1}b) and correspond to $\beta=\pm\alpha$. An error process can disrupt the stability of these states and induce transitions between them. By analogy with a particle in a double {well} potential, tunneling (or bit-flips) from one well to another is exponentially suppressed in the separation between the two wells (here defined as $|\alpha|^2)$, as long as the error process fulfills two criteria: it has to be local and sufficiently weak. {An error process is local if it} {transforms} {a state into neighboring states in phase space  \cite{Gottesman2001}. As an example, dominant errors such as photon loss, gain and dephasing are local}. Moreover, the {effective} error rate $\kappae$ must be weaker than the {confining rate} $\kappac=2|\alpha|^2\kappa_2$ \cite{supplement} inherited from the confining potential $V$, in order for the cat-qubit states to remain localized near the potential minima. The outstanding challenge to observe an exponential increase in the bit-flip time is therefore to engineer $\kappac>\kappae$ for all dominant local error processes.


\begin{figure*}
\includegraphics[width=\textwidth]{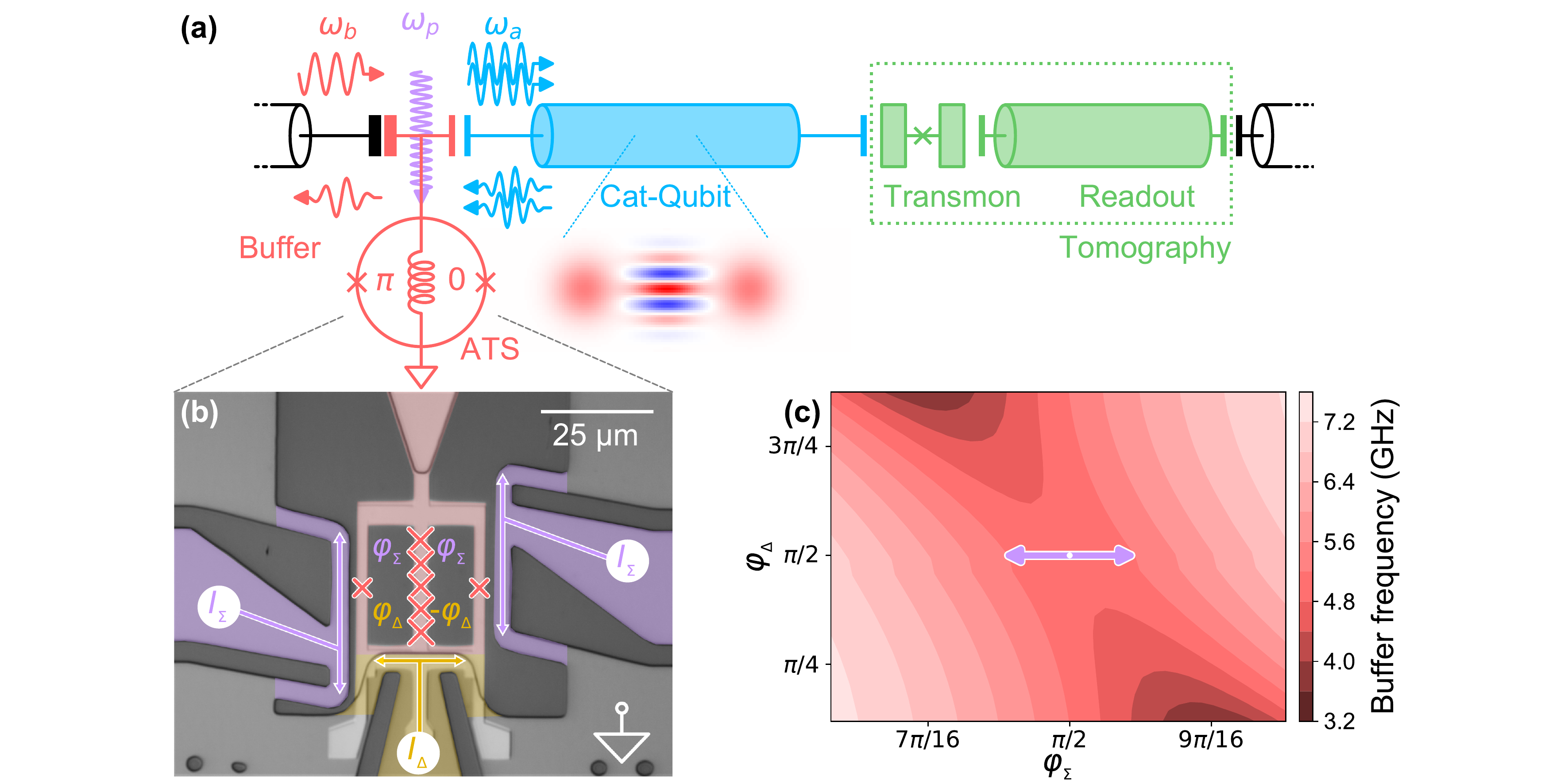}
\caption{\textbf{Circuit diagram and implementation} (\textbf{a}) The cat-qubit resonator (blue) is coupled on one end to a transmon qubit and a readout resonator (green) to measure its Wigner function, and on the other end to the buffer (red), a lumped element resonator connected to ground through a non-linear element coined the Asymmetrically Threaded SQUID (ATS). The ATS consists of a SQUID shunted by an inductance, forming two loops. Pumping the ATS at frequency $\omega_p=2\omega_a-\omega_b$ (purple arrow), where $\omega_{a,b}$ are the cat-qubit and buffer frequencies, mediates the exchange of two photons of the cat-qubit (blue arrows) with one photon of the buffer (red arrows) (\textbf{b}) {False color} optical image of the ATS. The shunt inductance is made of an array of 5 Josephson junctions ({marked by} large red crosses). The left and right flux lines (purple) are connected to the same input through an on-chip hybrid (not represented). They carry the radio-frequency pump and the DC current $I_{_\Sigma}$, which thread both loops with flux $\varphi_{_\Sigma}$. The bottom flux line (yellow) carries current $I_{_\Delta}$ and threads each loop with flux $\pm\varphi_{_\Delta}$. Combining these two controls, we bias the ATS at the $\pi/0$ asymmetric {DC} working point. (\textbf{c})~Measured buffer frequency (color) as a function of $\varphi_{_\Sigma}$ (x-axis) and $\varphi_{_\Delta}$ (y-axis), around the working point $\varphi_{_\Sigma}, \varphi_{_\Delta}= \pi/2, \pi/2$ (white dot). As expected, for $\varphi_{_\Sigma}=\pi/2$ (open SQUID), the buffer frequency does not depend on $\varphi_{_\Delta}$. We operate the ATS by modulating the flux along the orthogonal direction $\varphi_{_\Sigma}$ (purple arrow). From this {measurement}, we extract all the ATS parameters \cite{supplement}.}
\label{fig2}
\end{figure*}

Two-photon exchange between a resonator and its environment does not occur spontaneously. Instead, it is synthesized by engineering an interaction that exchanges pairs of photons of the cat-qubit resonator with one photon of an intentionally lossy mode referred to as the buffer \cite{Leghtas2015}. The interaction Hamiltonian takes the form
\begin{equation}
\label{eq:g2}
    \bm{H}_i/\hbar = g_2\ba^{\dag 2}\bm{b}+g_2^*\ba^2\bm{b}^\dag\,,
\end{equation}
where $\bm{b}$ is the annihilation operator of the buffer and $g_2$ is the interaction strength. 
Adding a resonant drive on the buffer, we recover \eqref{eq:H2} with
$\kappa_2\approx{4|g_2|^2}/{\kappa_b}$ and $\alpha^2 = -{\epsilon_d}/{g_2^*}$, 
where $\epsilon_d$ is the drive amplitude and $\kappa_b$ is the buffer energy decay rate, {engineered to be larger than $g_2$} \cite{Carmichael2007, Leghtas2015}.
Conveniently, the separation $|\alpha|^2$ between the cat-qubit states is readily tunable \textit{in situ} since it is proportional to the buffer drive amplitude.

{We implement our cat-qubit in a circuit quantum electrodynamics architecture described in Fig~\ref{fig2}a {operated at 10 mK}. It consists of a sputtered niobium film on a silicon substrate patterned into coplanar waveguide resonators. The cat-qubit mode resonates at $\omega_a/2\pi = 8.0381$~GHz, has a single photon lifetime $T_1 = 3.0~\mu$s and is probed through a transmon qubit coupled to a readout resonator followed by a parametric amplifier. {At the flux operating point,} the buffer mode resonates at $\omega_b/2\pi=4.8336$~GHz and has an energy decay rate $\kappa_b/2\pi = 13$~MHz.}

\begin{figure*}
\includegraphics[width=\textwidth]{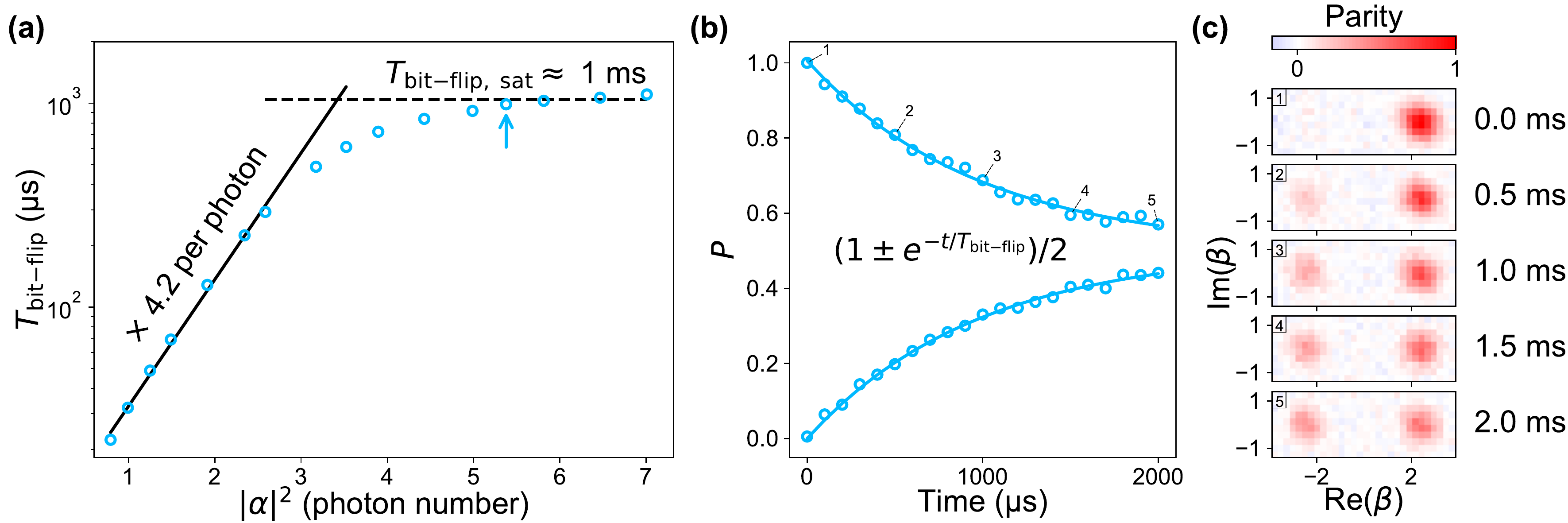}
\caption{\textbf{Exponential increase of the bit-flip time with the cat size.} (\textbf{a}) The bit-flip time (y-axis) is measured (open circles) as a function of the cat size {defined as} $|\alpha|^2$ (x-axis). Up to $|\alpha|^2\approx 3.5$, $T_\text{bit-flip}$ undergoes an exponential increase to $\approx 0.8$~ms, rising by a factor of 4.2 per added photon (solid line). The bit-flip time then saturates (dashed line is a guide for the eye) for $|\alpha|^2\ge 5$ at $1~$ms, a factor of 300 larger than the cat-qubit resonator lifetime $T_1$ in the absence of the pump and drive. {Each circle is obtained from measurements such as in (b) for the circle indicated by the blue arrow}. (\textbf{b}) The cat-qubit is initialized in $\kzero$, for a cat size $|\alpha|^2 = 5.4$. After applying the pump and drive for a variable duration (x-axis), the population $P$ (y-axis) of $\kzero$ (top curve) and $\kone$ (bottom curve) is measured. The data (open circles) are fitted to decaying exponential functions (solid lines) from which we extract the bit-flip time. (\textbf{c}) Each panel displays the measured Wigner function of the cat-qubit after a pump and drive duration indicated on the right of each plot. {Labels 1-5 mark the correspondence with  (b)}. The cat-qubit is initialized in $\kzero$ (top panel) and over a millisecond timescale, the population escapes towards $\kone$ (lower panels). The two-photon dissipation ensures that the cat-qubit resonator state remains entirely in the steady state manifold spanned by $\kzero$ and $\kone$.}
\label{fig4}
\end{figure*}

It is a technical challenge to engineer the interaction \eqref{eq:g2} without inducing spurious effects which are detrimental for the protection of quantum information. Examples of such effects are induced relaxation \cite{Sank2016, Gao2018}, escape to unconfined states \cite{LescannePRApp2019} and quasiparticle generation \cite{Wang2014}. To mitigate these effects, the interaction \eqref{eq:g2} is induced by a {novel} non-linear dipole: the Asymmetrically Threaded SQUID (ATS, Fig~\ref{fig2}b). The ATS consists of a symmetric SQUID (Superconducting Quantum Interference Device) shunted in its center by a large inductance, {thus forming two loops. Here the inductance} is built from an array of 5 Josephson junctions. The ATS mediates an interaction of the form {$U=-2E_{J}\cos(\varphi_{_\Sigma})\cos(\bvarphi+\varphi_{_\Delta})$, where $E_J$ is the Josephson energy of the SQUID junctions, $\bvarphi$ is the phase across the dipole, and $2\varphi_{_\Sigma,_\Delta}$ are the sum and differences of flux threading the two loops \cite{supplement}. We bias the ATS at $\varphi_\Sigma = \varphi_\Delta = \pi/2$, or equivalently, we thread the left and right loops with flux $\pi$ and $0$, respectively. In addition, we drive the sum port with a radio-frequency flux pump $\epsilon(t)$. At this bias point $U=-2E_{J}\sin(\epsilon(t))\sin(\bvarphi)$. The ATS is coupled to the buffer and cat-qubit, so that $\bvarphi$ is a linear combination of $\ba,\ba^\dag,\bb,\bb^\dag$, and $\sin(\bvarphi)$ contains only odd powers of these operators. The desired interaction \eqref{eq:g2} is present in the expansion of $\sin(\bvarphi)$, and is resonantly selected by a flux pump frequency $\omega_p=2\omega_a-\omega_b$ \cite{Vrajitoarea2018}}. {In contrast with previous strategies \cite{Leghtas2015, Touzard2018a}}, the ATS mediates a pristine two-photon coupling, since \eqref{eq:g2} is the only leading order non-rotating term, the presence of the inductive shunt prevents instabilities \cite{VerneyPRApp2019}, and the device operates at a first order flux insensitive point (Fig~\ref{fig2}c). {These features are key in order not to introduce inherent error processes that cannot be corrected by two-photon dissipation.}

The root advantage of the cat-qubit is that its computational states $\kzero$ and $\kone$ can be made arbitrarily long-lived simply by increasing the cat size $|\alpha|^2$, provided that $\kappac>\kappae$. In this experiment, the dominant error is due to energy decay so that $\kappae/2\pi=(2\pi T_1)^{-1}=53$~kHz \cite{supplement}, and $\kappac=2|\alpha|^2\kappa_2$ with {a measured} $\kappa_2/2\pi = 40$~kHz {(from which we infer $g_2/2\pi = 360 $~kHz)}. Hence, we enter the regime $\kappac>\kappae$ as soon as $|\alpha|^2 > 0.6$. We have measured that for each added photon in the cat-qubit state, the bit-flip time is multiplied by 4.2. This exponential scaling persists up to $|\alpha|^2\approx 3.5$, and the bit-flip time saturates for $|\alpha|^2\ge 5$ at 1 ms, a 300-fold improvement over the resonator intrinsic lifetime (see Fig.~\ref{fig4}). We expect a saturation {when the corrected bit flip rate reaches the rate of} residual errors which are not correctable, such as non-local errors. In the present experiment, we attribute this saturation to the coupling with the transmon employed for the resonator tomography \cite{supplement}, which has a thermal occupation of $1\%$, a lifetime $T_{1,q}=5~\mu$s and is dispersively coupled to the cat-qubit resonator with a rate $\chi/2\pi = 720~$kHz. Over a timescale in the millisecond range, the transmon acquires a thermal excitation {that} shifts the cat-qubit resonator frequency by $\chi$. This triggers a rotation of the resonator states which overcomes the confining potential since in this experiment $\chi \gg \kappac/2$ \cite{supplement} (note that tomography protocols compatible with smaller values of $\chi$ {have been recently demonstrated} \cite{Touzard2018b, Campagne2019}). During an average time $T_{1,q}$, the resonator states acquire an angle of order $\chi T_{1,q} \gg 2\pi$. When the transmon excitation decays, the rotation stops and the two-photon dissipation brings the resonator state back into the cat-qubit computational basis. By virtue of the dissipative nature of the protection mechanism, this process may result in a bit-flip but does not cause any leakage.

\begin{figure*}
\includegraphics[width=\textwidth]{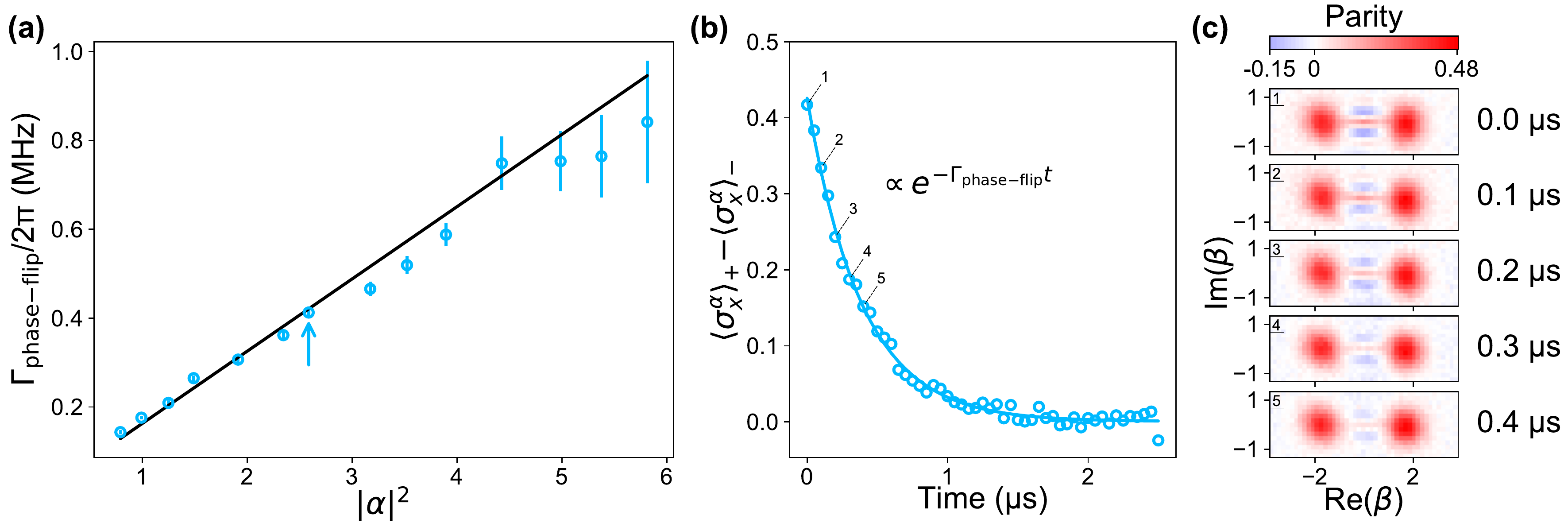}
\caption{\textbf{Linear increase of the phase-flip rate with the cat size}.
(\textbf{a}) The phase-flip rate (y-axis) is measured as a function of the cat size $|\alpha|^2$. The data (open circles) follow a linear trend (solid line) as expected for the decay rate of a Schr\"odinger cat coherence $\Gamma_\text{phase-flip}=2|\alpha|^2/T_{1,\mathrm{eff}}$. We measure $T_{1,\mathrm{eff}}=2.0~\mu$s, comparable to the intrinsic resonator lifetime of $3.0~\mu$s. {Each circle is obtained from measurements such as in (b) for the circle indicated by the blue arrow}. (\textbf{b}) The cat-qubit is prepared in the initial states $\ket{\pm}_{\alpha}$, for a cat size $|\alpha|^2 = 2.6$. After applying the pump and drive for a variable duration (x-axis), $\braket{\sigma_x^{\alpha}}_\pm$ is measured for each initial state and the difference is represented on the y-axis. The $X$ Pauli operator of the cat-qubit $\sigma_x^{\alpha}$ corresponds to the photon number parity. The data (open circles) are fitted to a decaying exponential (solid line) from which we extract the phase-flip rate. (\textbf{c}) Each panel displays the measured Wigner function of the cat-qubit after a pump and drive duration indicated on the right of each plot. {Labels 1-5 mark the correspondence with  (b)}. The cat-qubit is initialized in the $\ket{+}_{\alpha}$ state and the positive and negative fringes demonstrate the quantum nature of this initial state (top panel). The fringe contrast is reduced by single photon loss which mixes $\ket{+}_{\alpha}$ with $\ket{-}_{\alpha}$.}
\label{fig3}
\end{figure*}
%

Schr\"odinger cat states like $\ket{\pm}_\alpha$ living in a resonator with a lifetime $T_1$, lose their coherence at a rate $2|\alpha|^2/T_1$ \cite{Raimond2006}. In the cat-qubit paradigm, this translates into a phase-flip rate which increases linearly with the cat size $|\alpha|^2$. In addition, our cat-qubit undergoes a flux pump, a drive and non-linear interactions, which could further increase the phase-flip rate. We measure the phase-flip rate for increasing $|\alpha|^2$ and confirm a linear scaling (Fig.~\ref{fig3}a). Moving towards three dimensional cavities and engineering ever-improving non-linear interactions should decrease the phase-flip rate below a threshold where a line repetition code can actively correct remaining errors \cite{Guillaud2019}.

In conclusion, we have observed the exponential {decrease of the bit-flip rate} between our cat-qubit states $\kzero$ and $\kone$, as a function of their separation in phase space, while only linearly increasing their phase-flip rate. {Such an exponential scaling is necessary to bridge the gap between the modest performance of quantum hardware and the exquisite performance needed for quantum computation \cite{Fowler2012}}. This was made possible by inventing a Josephson circuit which mediates a pristine non-linear coupling between our cat-qubit mode and its environment. {Further improving the lifetime of the cavity to the state of the art of a millisecond \cite{Reagor2016} and a cat size of $|\alpha|^2\approx 5$ (resp. 10) should lead to a bit-flip time of $\approx 1$ second (resp. 0.5 hour), and a phase-flip time of $\approx 100~\mu$s (resp. 50~$\mu$s).} With such a long bit-flip time, the entire effort of active QEC will be focused on correcting the only significant error: phase-flips. In addition, conditional rotations in the 2D phase space of our cat-qubit form a universal set of gates, thus bypassing the need for magic states. These features suggest a significant reduction in hardware overhead for QEC \cite{Guillaud2019}.

%


\paragraph{Acknowledgements}
The authors acknowledge fruitful discussions with Pierre Rouchon and Clarke Smith. ZL acknowledges support from ANR project ENDURANCE, and EMERGENCES grant ENDURANCE of Ville de Paris.  AS acknowledges support from ANR project HAMROQS. The devices were fabricated within the consortium Salle Blanche Paris Centre. This work has been supported by the Paris \^Ile-de-France Region in the framework of DIM SIRTEQ.

{\paragraph{Author contributions}
RL designed, fabricated and measured the device, and analyzed the data. RL and ZL conceived the ATS element with help from BH and TP. RL and ZL wrote the paper with input from all authors. MV fabricated the parametric amplifier. TK and MD provided experimental support. AS and MM provided theory support. ZL managed the project. All authors contributed to extensive discussions of the results.}

\bibliographystyle{ieeetr}

\newpage

\clearpage

\onecolumngrid

\section{Supplementary materials}

\beginsupplement
\renewcommand{\bibnumfmt}[1]{[S#1]}
\renewcommand{\citenumfont}[1]{S#1}

\subsection{Full device and wiring}
\begin{figure}[!ht]
\includegraphics[width={510pt}]{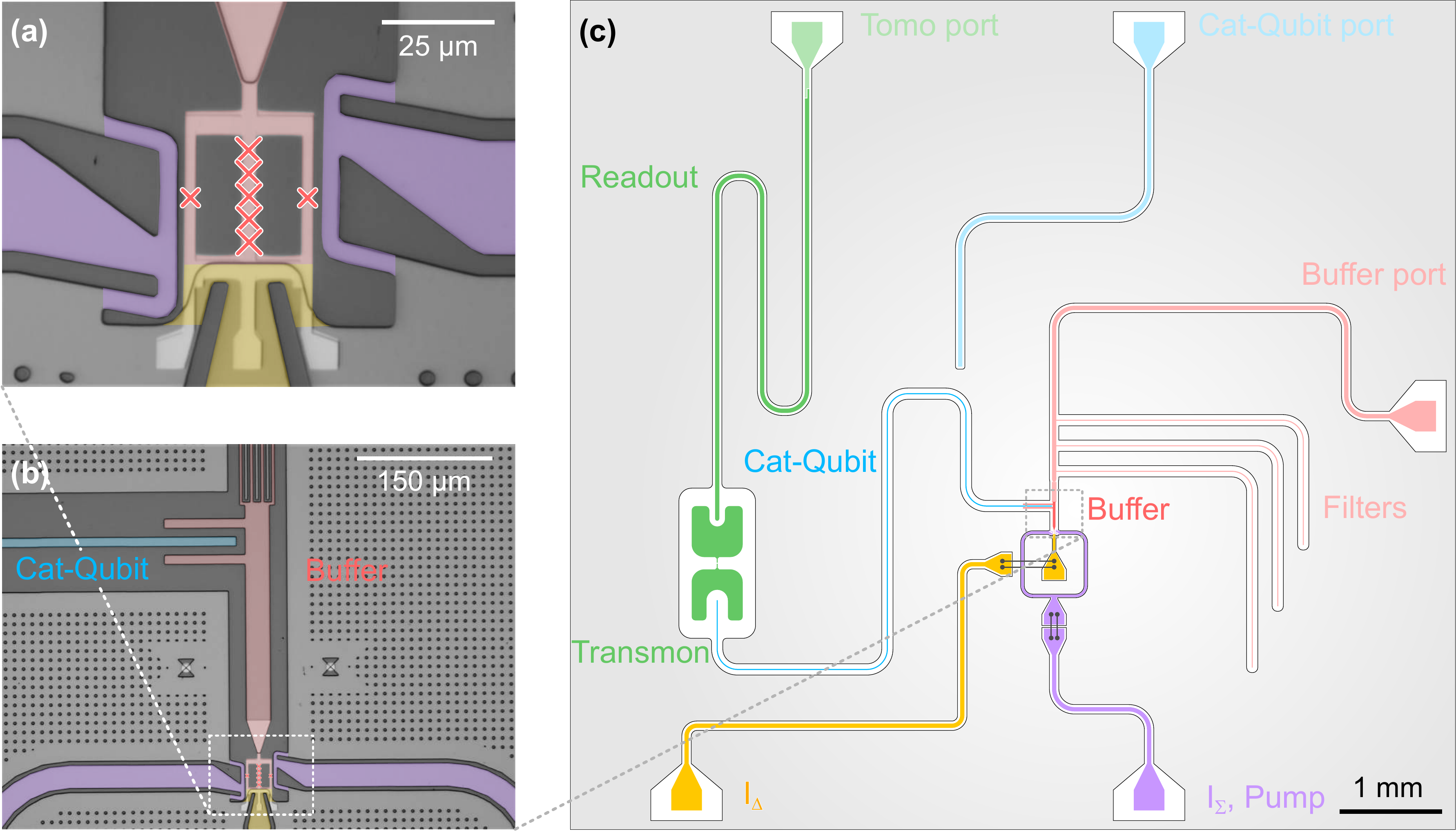}
\caption{\textbf{Full device layout.} (\textbf{a}) {False color} optical image of the ATS. Note that the 5 junction symbols are separated for clarity, the actual junctions are much closer and centered in the middle of the arm. (\textbf{b}) {False color} optical image of the buffer. The buffer (red) is strongly coupled to its transmission line via an interdigitated capacitor (top). It is also capacitively coupled to the cat-qubit resonator (blue). This is actually a picture of a twin sample where  this coupling was smaller. In panel (c), the real size of the coupling capacitor is shown. (\textbf{c}) Full device layout. The cat-qubit resonator is coupled on its other side to a transmon qubit, itself coupled to a readout resonator which together enable to perform the cat-qubit tomography. After the interdigitated capacitor, the buffer input is filtered via three $\lambda/4$-stub filters. These stop-band filters are centered at the cat-qubit resonance frequency to mitigate its direct coupling to the input line of the buffer \cite{Reed2010_supp} (b). The on-chip hybrid along the pump path (purple), equally splits the pump tone to RF-flux bias the ATS with the right symmetry. The black lines linking two dots are a schematic representation of the crucial wirebonds of the device. {The wirebonds linking the pump input to the on-chip hybrid where implemented to reduce the area of the loop delimited by the center conductor and the ground plane, leading to a reduced sensitivity to flux noise}.
}
\label{fig_device}
\end{figure}

\begin{figure}[!ht]
\includegraphics[width={350pt}]{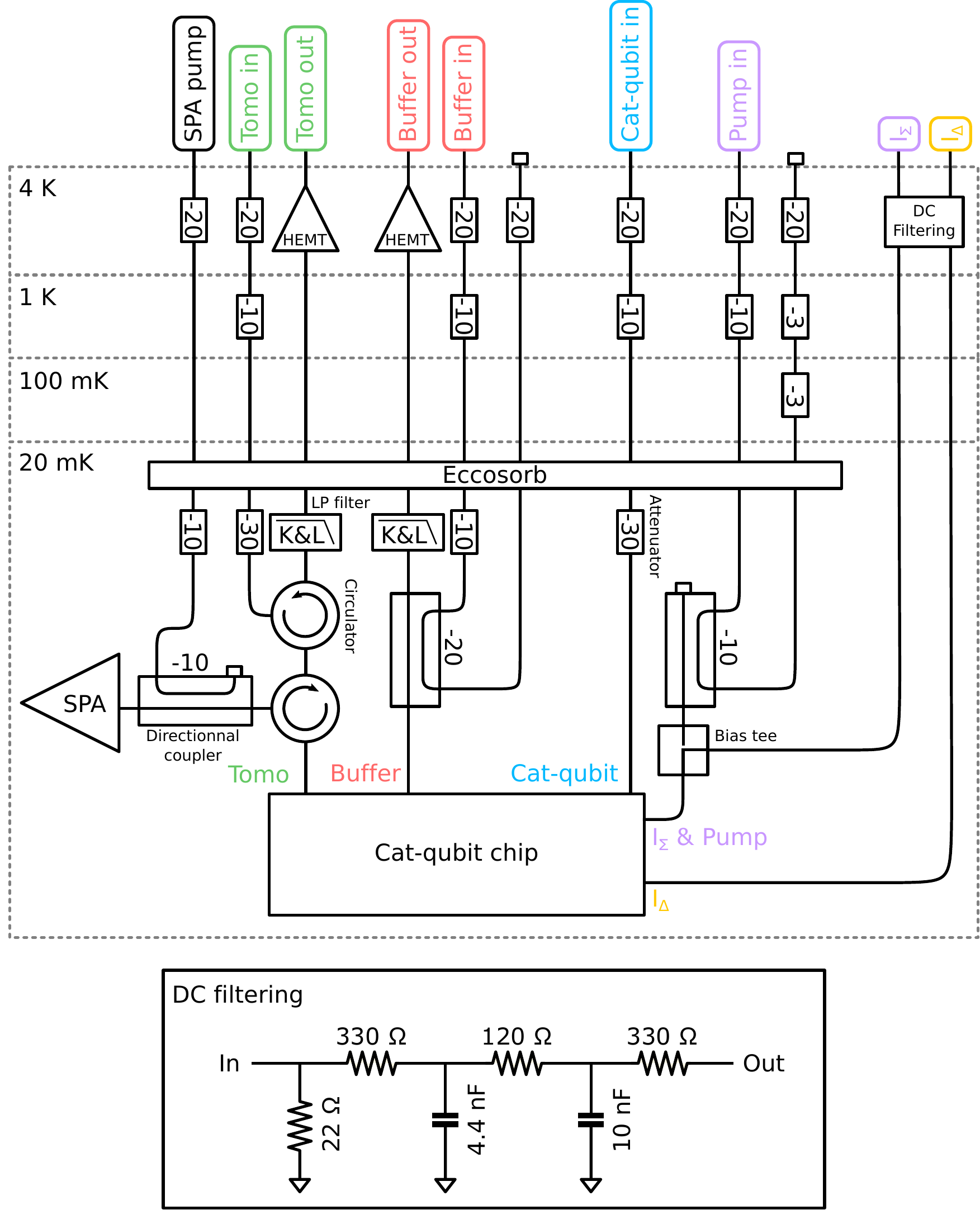}
\caption{\textbf{RF and DC wiring of the dilution refrigerator.} Note that the pump and drive tones are attenuated at the base plate via directional couplers so that the attenuated power is dissipated at higher fridge stages, far from the sample. {The $I_\Sigma$ DC current and the RF pump signal are combined at 20~mK with a bias-tee. We have used a homemade ``Snail Parametric Amplifier'' (SPA) \cite{Frattini2018_supp}.}
}
\label{fig_cabling}
\end{figure}

The circuit {consists of a sputtered niobium film} with a thickness of 120 nm deposited on a 280 $\mathrm{\mu}$m-thick wafer of intrinsic silicon. The main circuit is etched after an optical lithography step. The Josephson junctions are made of evaporated aluminum through a PMMA/MAA resist mask patterned in a distinct e-beam lithography step. A single Dolan bridge is used to make the small junctions of the ATS and of the transmon, and a series of 3 Dolan bridges delimit the 5-junction array which serves as the ATS inductor. The full device layout and the experiment wiring are displayed in Figs.~\ref{fig_device}, \ref{fig_cabling}.

\subsection{Hamiltonian derivation}
In this section, we derive the potential energy of the ATS dipole alone, and then calculate the full system Hamiltonian when the ATS mediates a non-linear coupling between the buffer and cat-qubit modes.

\begin{figure}[!ht]
\includegraphics[width=\textwidth]{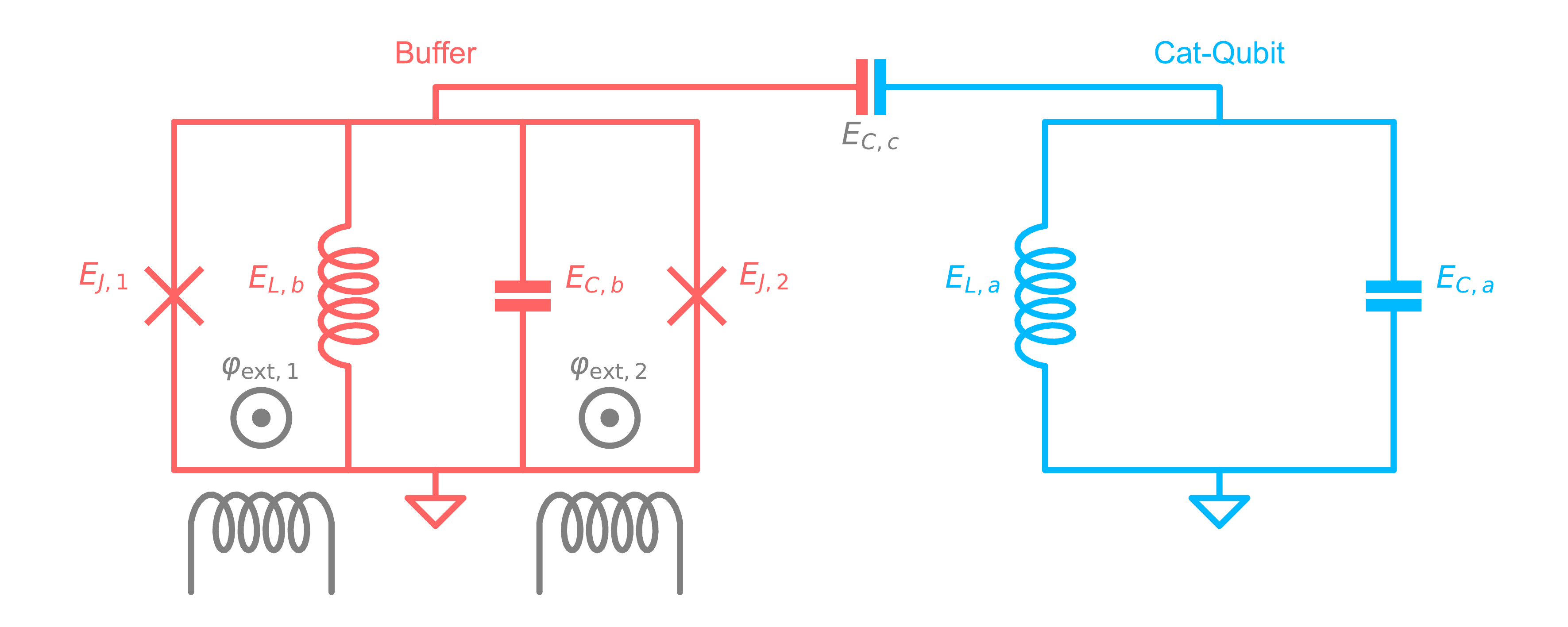}
\caption{Equivalent circuit diagram. The cat-qubit (blue), a linear resonator, is capacitively coupled to the buffer (red). One recovers the circuit of Fig.~\ref{fig2} by replacing the buffer inductance with a 5-junction array and by setting $\varphi_{_{\Sigma}}=(\varphi_{\mathrm{ext},1}+\varphi_{\mathrm{ext},2})/2$ and $\varphi_{_{\Delta}}=(\varphi_{\mathrm{ext},1}-\varphi_{\mathrm{ext},2})/2$. Not shown here: the buffer is capacitively coupled to a transmission line, the cat-qubit resonator is coupled to a transmon qubit}
\label{fig_circuit}
\end{figure}

\subsubsection{The potential energy of the ATS dipole element}

Let us first derive the potential energy of the ATS element alone. Its equivalent circuit is represented in red in Fig.~\ref{fig_circuit} and the phase across its inductor $\varphi$ is the only degree of freedom {(here we assume that the coupling capacitor to the cat-qubit mode is replaced by an open circuit)}. The potential energy of the ATS reads
\begin{equation}
    U(\varphi)=\frac{1}{2}E_{L,b}\varphi^2-E_{J,1}\cos(\varphi+\varphi_{\mathrm{ext},1})-E_{J,2}\cos(\varphi-\varphi_{\mathrm{ext},2})\;.
\end{equation}
Due to fabrication imperfections, the ATS junctions are not symmetric. We introduce $E_J$ and $\Delta E_J$ such that $E_{J,1}=E_J+\Delta E_J$ and  $E_{J,2}=E_J-\Delta E_J$. We obtain
\begin{equation}
    U(\varphi)=\frac{1}{2}E_{L,b}\varphi^2-2E_{J}\cos(\varphi_{_\Sigma})\cos(\varphi+\varphi_{_\Delta})+2\Delta E_J\sin(\varphi_{_\Sigma})\sin(\varphi+\varphi_{_\Delta})\;,
\end{equation}
{where $\varphi_{_{\Sigma}}=(\varphi_{\mathrm{ext},1}+\varphi_{\mathrm{ext},2})/2$ and $\varphi_{_{\Delta}}=(\varphi_{\mathrm{ext},1}-\varphi_{\mathrm{ext},2})/2$.} We DC bias the ATS at the asymmetric flux bias point $\varphi_{_\Sigma} = \varphi_{_\Delta}=\pi/2$ (this is a saddle point of the buffer frequency map). In addition, an RF flux bias on $\varphi_{_{\Sigma}}$ is applied, so that
\begin{eqnarray}
    \varphi_{_\Sigma}&=&\pi/2+\epsilon(t)\;,\;\text{with } \epsilon(t)=\epsilon_0\cos(\omega_p t)\\
    \varphi_{_\Delta}&=&\pi/2 \;.
\end{eqnarray}
The time-dependent potential at first order in $\epsilon(t)$ then reads
\begin{equation}
    U(\varphi)=\frac{1}{2}E_{L,b}\varphi^2-2E_{J}\epsilon(t)\sin(\varphi)+2\Delta E_{J}\cos(\varphi)\;.
\end{equation}
This potential is an unbounded function of $\varphi$, which prevents the system from escaping towards higher energy states in the presence of the pump \cite{VerneyPRApp2019_supp, LescannePRApp2019_supp}. In practice, with the 5-junction array replacing the inductance, the confining part of the potential is replaced by $5E_{J,L}\cos(\varphi/5)$ where $E_{J,L}=5E_{L,b}$ is the Josephson energy of each individual junction of the array. This potential is no longer unbounded, however the bound is high enough ($5E_{J,L}\gg 2E_{J}\epsilon_0$) for our pump power regime.

In the ideal case ($\Delta E_J=0$), this potential only produces odd powers of $\varphi$ from the sine non-linearity. A small asymmetry of the junctions produces small even powers of $\varphi$, leading to parasitic Kerr non-linearities. Typically $|\Delta E_J/E_J| \approx 10\%$. In the following we assume $\Delta E_J = 0$ for simplicity. 

\subsubsection{The coupled buffer and cat-qubit resonators}
We now consider the buffer and cat-qubit modes, and their coupling through the ATS dipole element. The full Hamiltonian reads
\begin{eqnarray}
    \bm{H}&=&\hbar\omega_{a,0}\bm{a}^\dag\bm{a}+\hbar\omega_{b,0}\bm{b}^\dag\bm{b}-2E_{J}\epsilon(t)\sin\left(\bm{\varphi}_b+\bm{\varphi}_a\right)\\
    \text{with }\bm{\varphi}_a &=& \varphi_{a}(\bm{a}+\bm{a}^\dag{})\,,\;
    \bm{\varphi}_b = \varphi_{b}(\bm{b}+\bm{b}^\dag{})
\end{eqnarray}
where $\ba/\bb$ are the annihilation operators of the cat-qubit and buffer modes, $\omega_{a/b,0}$ their resonant frequencies, and $\varphi_{a/b}$ their zero point phase fluctuations across the ATS dipole. Due to the circuit geometry, we expect $\varphi_{b}\gg \varphi_{a}$.
When expanding the sine up to third order in $\bm{\varphi} = \bm{\varphi_b} + \bm{\varphi_a}$ we get
\begin{equation}
    \begin{aligned}
    \bm{H}&=\hbar\omega_{a,0}\bm{a}^\dag\bm{a}+\hbar\omega_{b,0}\bm{b}^\dag\bm{b}-2E_{J}\epsilon(t)\varphi_{b}(\bm{b}+\bm{b}^\dag{})-2E_{J}\epsilon(t)\varphi_{a}(\bm{a}+\bm{a}^\dag{})\\
    &\qquad+\frac{1}{3}E_{J}\epsilon(t)\left(\bm{\varphi}_b+\bm{\varphi}_a\right)^3
    \end{aligned}
\end{equation}
The first two terms of the expansion are drives at frequency $\omega_p$ on the buffer and cat-qubit respectively. They can be absorbed in the frame displaced by $\xi_{a}(t) = \xi_{a}e^{-i\omega_pt}$ and $\xi_{b}(t) = \xi_{b}e^{-i\omega_pt}$ for $\ba$ and $\bb$ respectively, where
\begin{equation}
    \xi_{a/b} \xrightarrow[t\gg1/\kappa_{a/b}]\; \frac{i(E_J/\hbar)\epsilon_0\varphi_{a/b}}{\kappa_{a/b}/2+i(\omega_{a/b,0}-\omega_p)}
\end{equation}
In this displaced frame, the Hamiltonian reads
\begin{eqnarray}
    \bm{H}_\text{disp}&=&\hbar\omega_{a,0}\bm{a}^\dag\bm{a}+\hbar\omega_{b,0}\bm{b}^\dag\bm{b}\nonumber\\
    &+&\frac{1}{3}E_{J}\epsilon(t)\Big(\varphi_{b}(\bm{b}+\bm{b}^\dag{}+\xi_{b}e^{-i\omega_pt}+\xi^*_{b}e^{i\omega_pt})+\varphi_{a}(\bm{a}+\bm{a}^\dag{}+\xi_{a}e^{-i\omega_pt}+\xi^*_{a}e^{i\omega_pt})\Big)^3
    \label{eq:Hbig}
\end{eqnarray}



In practice, the buffer mode is driven with an additional microwave drive at frequency $\omega_d$, not included here for simplicity. We place ourselves in the frame rotating at $(\omega_p+\omega_d)/2$ and $\omega_d$ for $\bm{a}$ and $\bm{b}$ respectively. In this rotated frame, the Hamiltonian reads
\begin{eqnarray*}
    \bm{H_\text{rot}}&=&\hbar\left(\omega_{a,0}-\frac{\omega_p+\omega_d}{2}\right)\bm{a}^\dag\bm{a}+\hbar\left(\omega_{b,0}-\omega_d\right)\bm{b}^\dag\bm{b}\nonumber \\
    &+&\frac{1}{3}E_{J}\epsilon(t)\Big(\varphi_b(\bm{b}e^{-i\omega_d t} +\bm{b}^\dag{}e^{i\omega_d t}+\xi_{b}e^{-i\omega_pt}+\xi^*_{b}e^{i\omega_pt}) +
    \varphi_{a}(\bm{a}e^{-i\frac{\omega_p+\omega_d}{2} t}+\bm{a}^\dag{}e^{i\frac{\omega_p+\omega_d}{2} t}+\xi_{a}e^{-i\omega_pt}+\xi^*_{a}e^{i\omega_pt})\Big)^3
\end{eqnarray*}
Performing the rotating wave approximation (RWA), we get
\begin{equation}
    \bm{H_\text{RWA}}/\hbar=\left(\omega_{a}-\frac{\omega_p+\omega_d}{2}\right)\bm{a}^\dag\bm{a}+\left(\omega_b-\omega_d\right)\bm{b}^\dag\bm{b}+
    g_2^*\ba^2\bm{b}^\dag+g_2\ba^{\dag 2}\bm{b}\;,
    \label{eqsup:g2}
\end{equation}
where the modes frequencies are AC-Stark shifted to $\omega_{a/b} = \omega_{a/b,0}-\Delta_{a/b}$ and 
\begin{equation}
\hbar\Delta_{a/b}= \frac{1}{3}E_J\varphi_{a/b}^2\left(\mathrm{Re}(\xi_b)\varphi_{b}+\mathrm{Re}(\xi_a)\varphi_{a}\right)
\end{equation}
with $\hbar g_2 = E_J\epsilon_0\varphi_a^2\varphi_b/2$. When we verify the frequency matching condition
$$\omega_d=\omega_b, \qquad \omega_p=2\omega_a-\omega_b\;,$$ we recover Eq.~\eqref{eq:g2}, which we recall here
\begin{equation*}
    \bm{H_i}/\hbar = g_2^*\ba^2\bm{b}^\dag+g_2\ba^{\dag 2}\bm{b}\;.
\end{equation*}

\subsection{Circuit parameters}
\begin{figure}[!ht]
\includegraphics[width={510pt}]{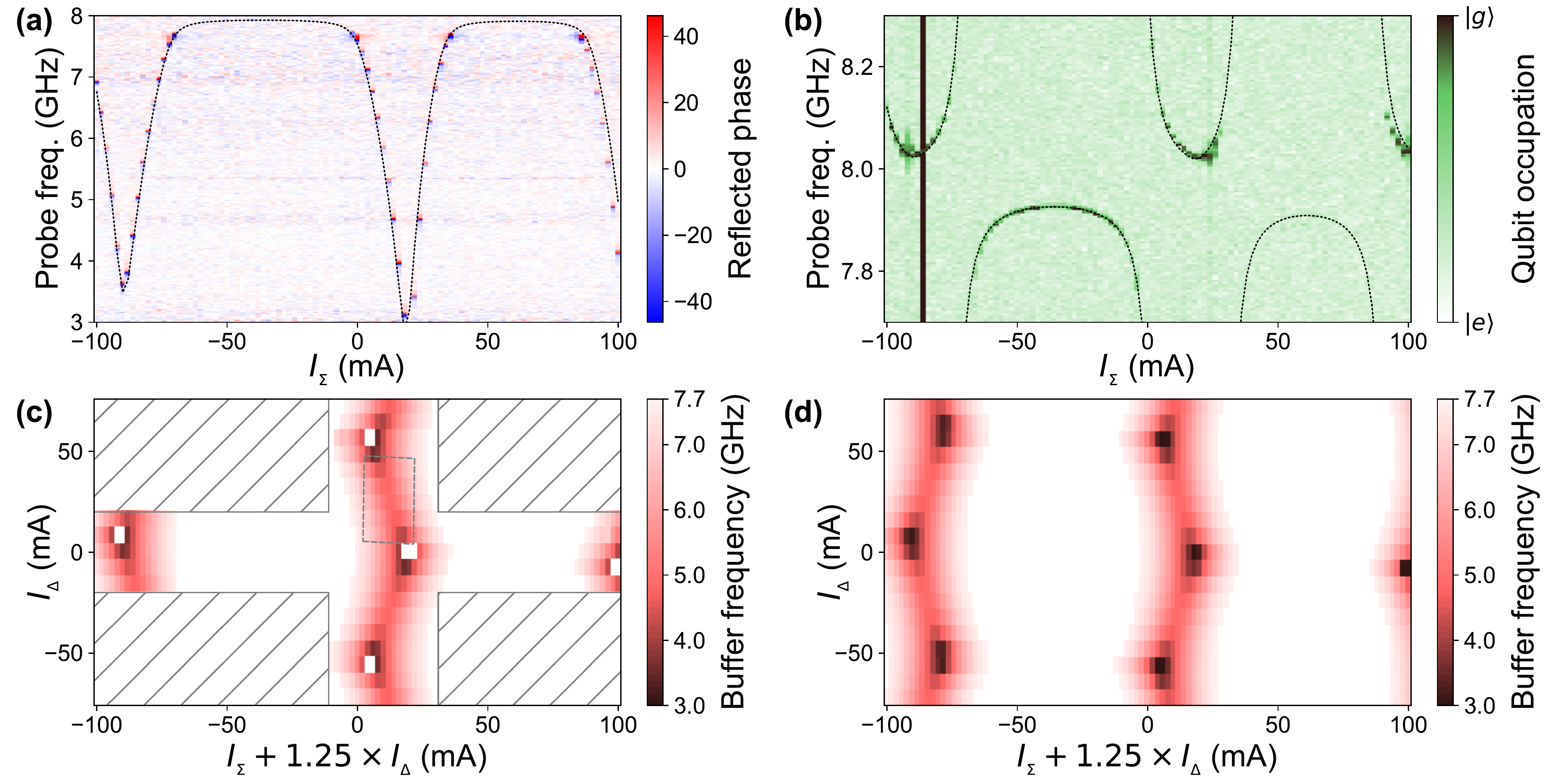}
\caption{\textbf{Flux dependence.} (\textbf{a}) Buffer spectroscopy. Phase of the reflected probe signal (colormap) on the buffer port as a function of $I_{_{\Sigma}}$ (x-axis) and probe frequency (y-axis). The buffer frequency follows an arch-like pattern typical of SQUID-based devices. The probing frequency range is limited to 3-8 GHz due to the 4-8 GHz circulator on the output line. The black dashed line represents the expected buffer/cat-qubit frequency for the best fitting parameter set. The slight aperiodicity may be explained by small asymmetries in the loop areas of the ATS. (\textbf{b}) Cat-qubit resonator two-tone spectroscopy. A continuous probe is applied on the cat-qubit resonator at various frequencies (y-axis) and a second tone attempts to $\pi$-pulse the qubit on resonance. When the probe populates the cat-qubit, the qubit shifts in frequency due to the cross-Kerr coupling, and is insensitive to the $\pi$-pulse. The resulting qubit state is plotted in color and we repeat the experiment for various values of $I_{_\Sigma}$ (x-axis). The black dashed line represents the expected buffer/cat-qubit frequency for the best fitting parameter set. The black vertical line corresponds to a flux bias were the buffer is at the qubit frequency resulting in a strong decrease in the qubit lifetime. (\textbf{c}) For each value of $I_{_{\Sigma}}$ (x-axis) and $I_{_{\Delta}}$ (y-axis), we extract the buffer frequency from a spectroscopy measurement (panel a) and report it in color (white is when the resonance frequency is beyond the measurement range). Contrary to what they were designed for, the two flux lines do not perfectly apply symmetric ($I_{_{\Sigma}}$) and antisymmetric ($I_{_{\Delta}}$) bias on the ATS. We compensate for this imbalance while taking the data by shifting the $I_{_{\Sigma}}$ span for each value of $I_{_{\Delta}}$ as indicated on the x-axis label. Data were only taken on the area outside the hatched regions to prevent the heating of the dilution refrigerator beyond a tolerable temperature. The grey dashed-rectangle corresponds to the flux range presented in Fig.~\ref{fig2} of the main text.
(\textbf{d}) Simulated flux dependence of the buffer mode for the best fitting parameter set.}
\label{fig_flux}
\end{figure}

Most of the circuit parameters can be readily deduced from standard circuit-QED measurements and are gathered in Table~\ref{table}. Here we explain the methodology we used to deduce the 6 dipole parameters of Fig.~\ref{fig_circuit} (see Table~\ref{table0}) and the mapping of $(I_{_\Sigma}, I_{_\Delta})$ to $(\varphi_{_\Sigma}, \varphi_{_\Delta})$. Independently of this mapping, the ATS saddle point is unambiguously found. At this flux point, $E_J\cos(\varphi_{_\Sigma})=0$,   and we directly measure $\omega_{a_0}$ and $\omega_{b,0}$. 
The energies $E_J$ and $E_{L,b}$ are computed from the Ambegaokar-Baratoff and the room temperature measurements of neighbouring test junction resistances. The general linear transformation mapping $(I_{_\Sigma}, I_{_\Delta})$ to $(\varphi_{_\Sigma}, \varphi_{_\Delta})$ is found by fitting the measured buffer frequency as a function of $(I_{_\Sigma}, I_{_\Delta})$ (see Fig.~\ref{fig_flux}c,d). The impedance $Z_a$ of the cat-qubit resonator is estimated from the aspect ratio of the coplanar waveguide geometry. The energy $E_{C,c}$ is adjusted to match the measured anti-crossing of the buffer and cat-qubit mode when $I_{_\Sigma}$ is varied (see Fig.~\ref{fig_flux}b).

\begin{table}[!ht]
\begin{center}
\begin{tabular}{ccccc}
    \begin{tabular}[t]{| c | c |}
    \multicolumn{2}{c}{Cat-qubit mode}       \\ \hline
    $\omega_{a}/2\pi$& $8.03805\ \mathrm{GHz}$ \\ \hline
    $\omega_{a,0}/2\pi$& $8.0389\ \mathrm{GHz}$ \\ \hline
    $T_1$ & $3\ \mathrm{\mu s}$  \\ \hline
    $\kappa_a/2\pi$& $ 53\ \mathrm{kHz}$  \\ \hline
    $\chi_{aa}/2\pi$& $ -7\ \mathrm{kHz}$  \\ \hline
    \end{tabular}&
    \begin{tabular}[t]{| c | c |}
      \multicolumn{2}{c}{Buffer}         \\ \hline
    $\omega_{b}/2\pi$& $4.8336\ \mathrm{GHz}$ \\ \hline
    $\omega_{b,0}/2\pi$& $4.886\ \mathrm{GHz}$ \\ \hline
    $\kappa_b/2\pi$ & $13\ \mathrm{MHz}$  \\ \hline
    $\chi_{bb}/2\pi$& $ -32\ \mathrm{MHz}$  \\ \hline
    $\chi_{ba}/2\pi$& $ 0.79\ \mathrm{MHz}$  \\ \hline
    \end{tabular}&
    \begin{tabular}[t]{| c | c |}
    \multicolumn{2}{c}{Pump}       \\ \hline
    $\omega_p/2\pi$& $11.2425\ \mathrm{GHz}$ \\ \hline
    \end{tabular}&
    \begin{tabular}[t]{| c | c |}
    \multicolumn{2}{c}{Transmon}       \\ \hline
    $\omega_q/2\pi$& $4.4156\ \mathrm{GHz}$ \\ \hline
    $T_{1,q}$ & $5\ \mathrm{\mu s}$  \\ \hline
    $T_{2,q}$ & $8\ \mathrm{\mu s}$   \\ \hline
    $\chi_{qq}/2\pi$& $180\ \mathrm{MHz}$ \\ \hline
    $\chi_{qa}/2\pi$& $720\ \mathrm{kHz}$ \\ \hline
    $\pi/\chi_{qa}$& $0.69\ \mathrm{\mu s}$ \\ \hline
    \end{tabular}&
    \begin{tabular}[t]{| c | c |}
    \multicolumn{2}{c}{Readout}       \\ \hline
    $\omega_r/2\pi$& $6.4598\ \mathrm{GHz}$ \\ \hline
    ${\kappa_r}/2\pi$& $ 1.47\ \mathrm{MHz}$  \\ \hline
    \end{tabular}
    \tabularnewline
\end{tabular}
\caption{Measured system parameters at the ATS working point. The pump shifts the cat-qubit resonator and buffer frequencies. The frequencies in the absence of the pump are noted $\omega_{a/b,0}$ and those in its presence are denoted $\omega_{a/b}$. The Kerr couplings $\chi_{mn}$ enter the full Hamiltonian in the form $-\chi_{mn}m^\dag m n^\dag n$ when $m\ne n$ and $-\frac{\chi_{mm}}{2}{m^\dag}^2 m^2$, where $m, n$ denote the mode indices.}
\label{table}
\end{center}
\end{table}

\begin{table}[!ht]
\begin{center}
\begin{tabular}{cc}
    \begin{tabular}[t]{| c | c |}
    \multicolumn{2}{c}{Circuit parameters}       \\ \hline
    $\omega_{a,0}/2\pi$ & $8.0389\ \mathrm{GHz}$ \\ \hline
    $Z_a$& $90\ \Omega$ \\ \hline
    $\omega_{b,0}/2\pi$& $4.886\ \mathrm{GHz}$ \\ \hline
    $E_{C,c}/h$ & $720\ \mathrm{MHz}$ \\ \hline \hline
    $E_{L,b}/h$ & $45\ \mathrm{GHz}$ \\ \hline
    $E_{J}/h$ & $90\ \mathrm{GHz}$ \\ \hline
    \end{tabular}
    \begin{tabular}[t]{| c | c |}
    \multicolumn{2}{c}{Dipole parameters}       \\ \hline
    $E_{L,a}/h$& $96.6\ \mathrm{GHz}$ \\ \hline
    $E_{C,a}/h$& $92.7\ \mathrm{MHz}$ \\ \hline
    $E_{C,b}/h$& $73.5\ \mathrm{MHz}$ \\ \hline
    $E_{C,c}/h$& $720\ \mathrm{MHz}$ \\ \hline \hline
    $E_{L,b}/h$& $45\ \mathrm{GHz}$ \\ \hline 
    $E_{J}/h$ & $90\ \mathrm{GHz}$ \\ \hline
    \end{tabular}
    \tabularnewline
\end{tabular}
\caption{Measured and estimated circuit parameters (left), and their corresponding dipole energies (right).}
\label{table0}
\end{center}
\end{table}

\subsection{Semi-classical analysis}
In this section, we compute the semi-classical dynamics of the cat-qubit resonator in the presence of various imperfections (single photon loss and detuning). We gain insight into these dynamics by introducing a pseudo-potential function. For a more complete approach, we refer the reader to Refs.~\cite{CohenThesis_supp,Mirrahimi2019_supp}.
\subsubsection{Two-photon dynamics}
Under two-photon dissipation, the cat-qubit resonator state $\rho$ undergoes the following dynamics
\begin{equation}
\frac{d}{dt}\rho= \kappa_2 D[\bm{a}^2-\alpha^2]\rho\;,
\end{equation}
where the Lindblad operator $D$ is defined for any operator $\bm{O}$ as $D[\bm{O}]\rho = \bm{O}\rho\bm{O}^\dag - \frac{1}{2}\rho\bm{O}^\dag\bm{O}-\frac{1}{2}\bm{O}^\dag\bm{O}\rho $. Any combination of the states $\ket{0,1}_\alpha$ is a steady state of this dynamics. Moreover, these steady states are global attractors. To gain insight, we restrict this dynamics to the set of coherent states $\rho(t)=\ket{\beta(t)}\bra{\beta(t)}$, and introduce the pseudo-potential $V$ defined over the resonator phase space as $-\nabla V(\beta)=\frac{d\beta}{dt}$. This pseudo-potential depicts in which direction of the phase space a coherent state $\ket{\beta}$ evolves, and coherent steady states of the dynamics are the minima of $V$. Following ref.~\cite{Leghtas2015_supp}, we have
\begin{equation}
    \frac{d\beta}{dt} = -\kappa_2 \beta^*(\beta^2-\alpha^2)\;.
    \label{eq:velocity2}
\end{equation}
In the following we introduce $x=\re{\beta}$ and $y=\im{\beta}$ and we consider $\alpha$ real. Separating the real and imaginary part of equation \eqref{eq:velocity2}, we get
\begin{eqnarray}
    \frac{dx}{dt} &=& -\kappa_2\left(x^3 + xy^2-x\alpha^2\right)\nonumber\\
    \frac{dy}{dt} &=& -\kappa_2\left(y^3 + yx^2+y\alpha^2\right)\;.\nonumber
\end{eqnarray}
The velocity of a coherent state $\ket{\beta}$ in phase space is $(\frac{dx}{dt},\frac{dy}{dt})$ (see Fig.~\ref{fig_pot}a,\ref{fig_pot_delta}a). By integrating this velocity over space, we get the pseudo-potential
\begin{equation}
    V(x,y) =\kappa_2\left(\frac{1}{4}(x^4+y^4)+\frac{1}{2}x^2y^2-\alpha^2(x^2-y^2)\right)
    \label{eq:pseudopot2}
\end{equation}
depicted in Fig.~\ref{fig1}b of the main text. It has two minima in $-\alpha$ and $\alpha$. Analyzing the evolution of small deviations $\delta x$ and $\delta y$ around these minima, we find
\begin{eqnarray*}
    \frac{d}{dt}\delta x &=& -\kappac \delta x \\
    \frac{d}{dt}\delta y &=& -\kappac \delta y \;,
\end{eqnarray*}
where the confinement rate $\kappac$ is defined as
\begin{equation}
\kappac = 2\kappa_2 \alpha^2\;.
\label{eq:conf}
\end{equation}

This confinement pins down a computational state at each potential minimum, and protects the cat-qubit against errors. Next, we analyze the effect of errors on the cat-qubit resonator.
\subsubsection{Single photon loss}

When added, most Hamiltonian or dissipative mechanisms (such as detuning, single photon loss or gain, and dephasing) will perturb the system so that the two-dimensional cat-qubit space is no longer a steady-manifold of the overall dynamics. Instead, only one mixed state is a steady-state. However, this steady-state is exponentially (in $|\alpha|^2$) long to reach from the cat-qubit computational states $\kzero$ and $\kone$ \cite{CohenThesis_supp,Mirrahimi2019_supp}. We refer to these states as metastable states. We will now use the pseudo-potential representation to visulalize the main effects of single photon loss and detuning.

\begin{figure}[!ht]
\includegraphics[width={510pt}]{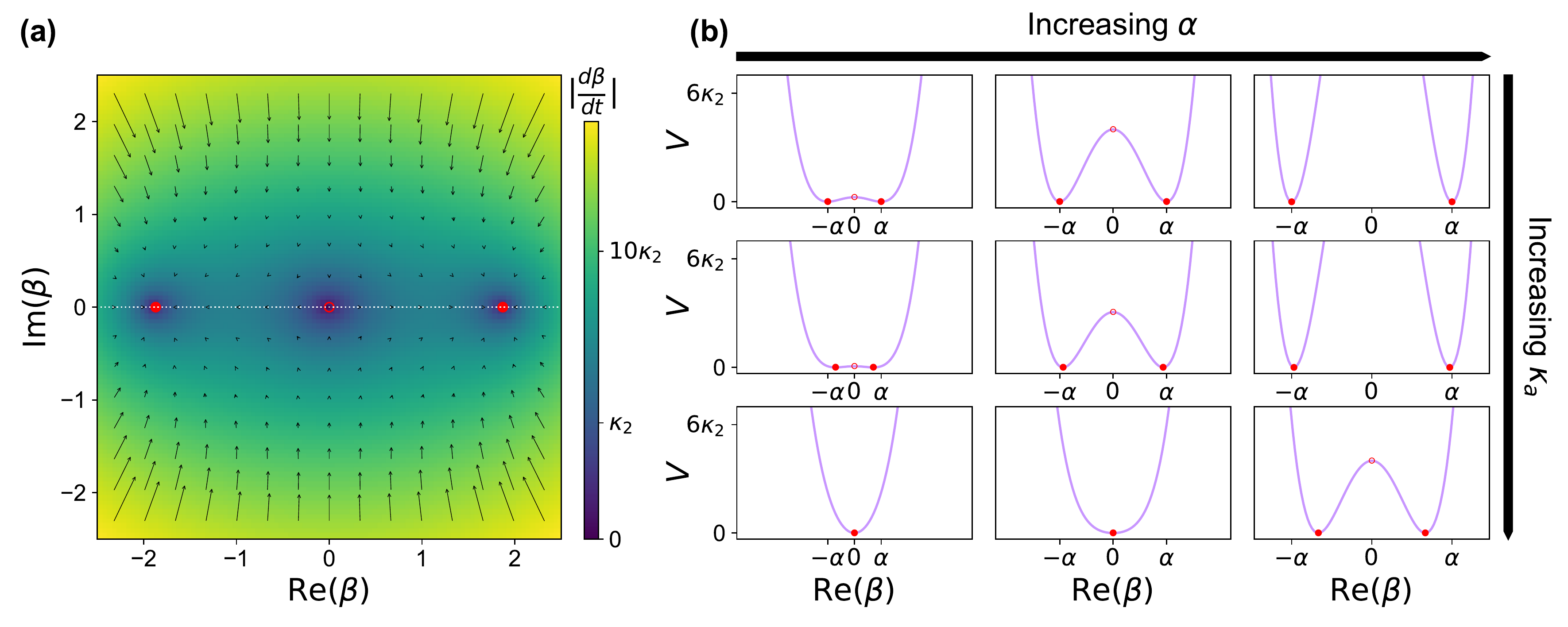}
\caption{\textbf{Dynamics under single photon loss.} \textbf{(a)} This colormap represents the magnitude of the velocity field \eqref{eq:velocity1} over phase space ($\beta$ plane) for $\kappa_a=\kappa_2$ and $\alpha=2$. The black arrows represent the velocity at various locations. The two stable/ one unstable steady-states are indicated with full/open red circles (the stability is infered from the direction of the arrows). The white dotted line represents the cut along which we plot the potential. \textbf{(b)} Cuts ($\re{\beta}=0$) of the potential \eqref{eq:pseudopot1} with $\alpha=(1,2,3)$ (left to right) and $\kappa_a= (0, \kappa_2, 10\kappa_2)$ (top to bottom). The top row represents the unperturbed potential and the steady states are $-\alpha$ and $\alpha$ (red circles). As we increase $\kappa_a$, the amplitude $\ainf$ of the metastable states (red circles) decreases until reaching 0. However for a given value of $\kappa_a$, we can always find a value of $\alpha$ to recover two metastable states (bottom row).
}
\label{fig_pot}
\end{figure}

Let us calculate $V$ in the presence of single photon loss at rate $\kappa_a$. The loss operator is $\bm{L}_1 = \sqrt{\kappa_a}\ba$ and the overall dynamics reads

\begin{equation}
\frac{d}{dt}\rho= \kappa_2 D[\bm{a}^2-\alpha^2]\rho + \kappa_a D[\bm{a}]\rho \,.
\label{eq:me_1phloss}
\end{equation}
Following the same computation as previously, we have
\begin{equation}
    \frac{d\beta}{dt} = -\kappa_2 \beta^*(\beta^2-\alpha^2) - \frac{1}{2}\kappa_a\beta \nonumber
\end{equation}
so that
\begin{eqnarray}
    \frac{dx}{dt} &=& -\kappa_2\left(x^3 + xy^2-x\alpha^2\right) -\frac{1}{2}\kappa_ax\nonumber\\
    \frac{dy}{dt} &=& -\kappa_2\left(y^3 + yx^2+y\alpha^2\right) -\frac{1}{2}\kappa_ay\,.
    \label{eq:velocity1}
\end{eqnarray}
This velocity field is represented in Fig.~\ref{fig_pot}a for $\kappa_a=\kappa_2$ and $\alpha=2$.
By integrating it over space we get
\begin{equation}
    V(x,y) =\kappa_2\left(\frac{1}{4}(x^4+y^4)+\frac{1}{2}x^2y^2-\alpha^2(x^2-y^2)\right)+\frac{1}{4}\kappa_a \left(x^2+y^2\right)\,.
    \label{eq:pseudopot1}
\end{equation}
Cuts that pass through the two minima are plotted in Fig.~\ref{fig_pot}b for various values of $\kappa_a$ and $\alpha$. 

The minima of $V$ are located in $\pm\ainf$ with
\begin{equation}
\ainf = 
\begin{cases} \sqrt{\alpha^2-\kun/(2\kde)} &\mbox{if } \alpha^2\ge \kun/(2\kde) \\
0 & \mbox{otherwise}
\end{cases}
\label{eq:ainf}
\end{equation}
In this semi-classical analysis, we find that two metastable states form when the error rate $\kappa_a$ is below the threshold
$$
\kappa_a < \kappac = 2|\alpha|^2\kappa_2\;.
$$

\subsubsection{Detuning}
\begin{figure}[!ht]
\includegraphics[width={510pt}]{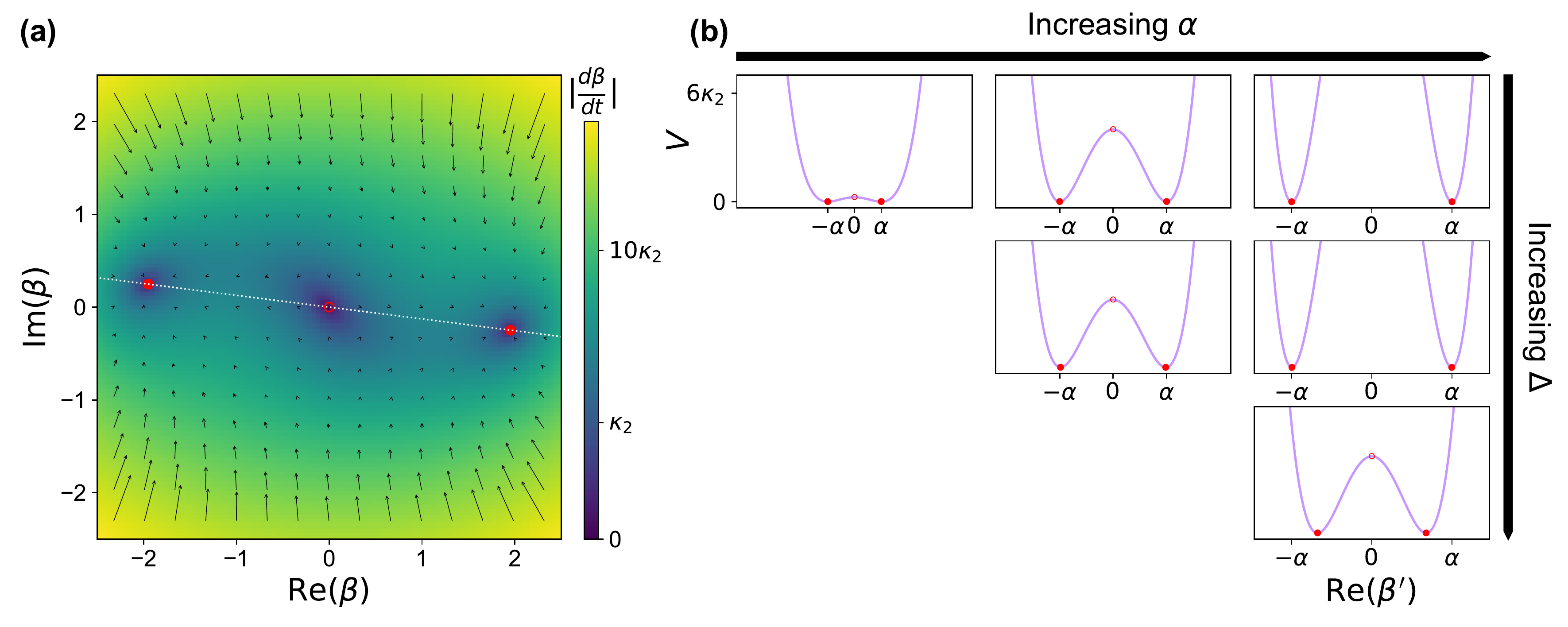}
\caption{\textbf{Dynamics under detuning.} \textbf{(a)} This colormap represents the magnitude of the velocity field \eqref{eq:velocitydet} over the phase space ($\beta$ plane) for $\Delta=\kappa_2$ and $\alpha=2$. The black arrows represent the velocity at various locations. The two stable/ one unstable steady-states are indicated with full/open red circles (the stability is infered from the knowledge of the direction of the arrows). The white dotted line represents the cut along which we represent the potential. \textbf{(b)} Cuts ($\im{\beta}=\lambda\re{\beta}$) of the potential \eqref{eq:pseudopotdet} with $\alpha=(1,2,3)$ (left to right) and $\Delta= (0, \kappa_2, 8\kappa_2)$ (top to bottom). The top row represents the unperturbed potential and the steady states are $-\alpha$ and $\alpha$ (red circles). As we increase $\Delta$, the amplitude $|\ainf|$ of the metastable states (red circles) decreases until reaching 0. However for a given value of $\Delta$, we can always find a value of $\alpha$ to recover two metastable states (bottom row).
}
\label{fig_pot_delta}
\end{figure}

 In the main text, we discussed the causes of the bit-flip time saturation and blamed the random frequency shifts of the cat-qubit resonator induced by the transmon thermal excitations. Let us see how a detuning $\Delta$ of the cat-qubit frequency affects the two-photon stabilization. In this case, we have
\begin{equation}
    \frac{d\beta}{dt} = -\kappa_2 \beta^*(\beta^2-\alpha^2)-i\Delta\beta
\end{equation}
so that
\begin{eqnarray}
    \frac{dx}{dt} &=& -\kappa_2\left(x^3 + xy^2-x\alpha^2\right) + \Delta y \nonumber\\
    \frac{dy}{dt} &=& -\kappa_2\left(y^3 + yx^2+y\alpha^2\right) - \Delta x \,.
    \label{eq:velocitydet}
\end{eqnarray}
Note that $\mathrm{rot}(\frac{dx}{dt}, \frac{dy}{dt}) = -2\Delta\ne 0$ so we cannot perform the spatial integration to find $V(x,y)$. We can obtain the steady states directly by analyzing the velocity field (Fig.~\ref{fig_pot_delta}a) but there exists a direction in phase-space parametrized by a real parameter $\lambda$ such that $y=\lambda x$ and $\frac{dy}{dt}=\lambda \frac{dx}{dt}$ along which the integration is meaningful. Plugging in this relation into \eqref{eq:velocitydet} we get the following condition on $\lambda$
\begin{eqnarray}
    \lambda^2\Delta +2\lambda\kde\alpha^2+\Delta=0 \nonumber\\
    \lambda = -\frac{\kappac}{2\Delta} + \sqrt{(\frac{\kappac}{2\Delta})^2-1}
\end{eqnarray}
with $\kappac=2\kde\alpha^2$. We have chosen the solution $\lambda$ which approaches 0 when $\Delta\rightarrow 0$ and for which the chosen direction crosses the steady states. Along this cut indexed by $\beta'$, we have 
\begin{equation}
    \frac{d\beta'}{dt} = \left(\sqrt{\left(\frac{\kappac}{2}\right)^2-\Delta^2}\right)\beta'-\kde\beta'^3 
\end{equation}
leading to 
\begin{equation}
    V(\beta') = -\frac{1}{2}\left(\sqrt{\left(\frac{\kappac}{2}\right)^2-\Delta^2}\right)\beta'^2+\frac{1}{4}\kde\beta'^4  
    \label{eq:pseudopotdet}
\end{equation}
that is plotted in Fig.~\ref{fig_pot_delta}b. There are two minima located along the direction $\beta'$ in 
\begin{equation}
|\ainf| = 
\begin{cases} \left(\alpha^4-\left(\frac{\Delta}{\kde}\right)^2\right)^\frac{1}{4} &\mbox{if } \Delta<\frac{\kappac}{2} \\
0 & \mbox{otherwise}
\end{cases} 
\label{eq:conf_det}
\end{equation}
In this semi-classical analysis, we find that two metastable states form provided $\Delta$ is below the threshold
$$
\Delta < \kappac/2 = |\alpha|^2\kappa_2\;.
$$
In our experiment, the detuning induced by a thermal photon entering the transmon is $\Delta/2\pi=\chi/2\pi=720\,\mathrm{kHz}$ which is larger than $\kappac/2/2\pi= 7\kde/2\pi\approx280\,\mathrm{kHz}$ for the largest $|\alpha|^2=7$.

\begin{figure}[!ht]
\includegraphics[width={246pt}]{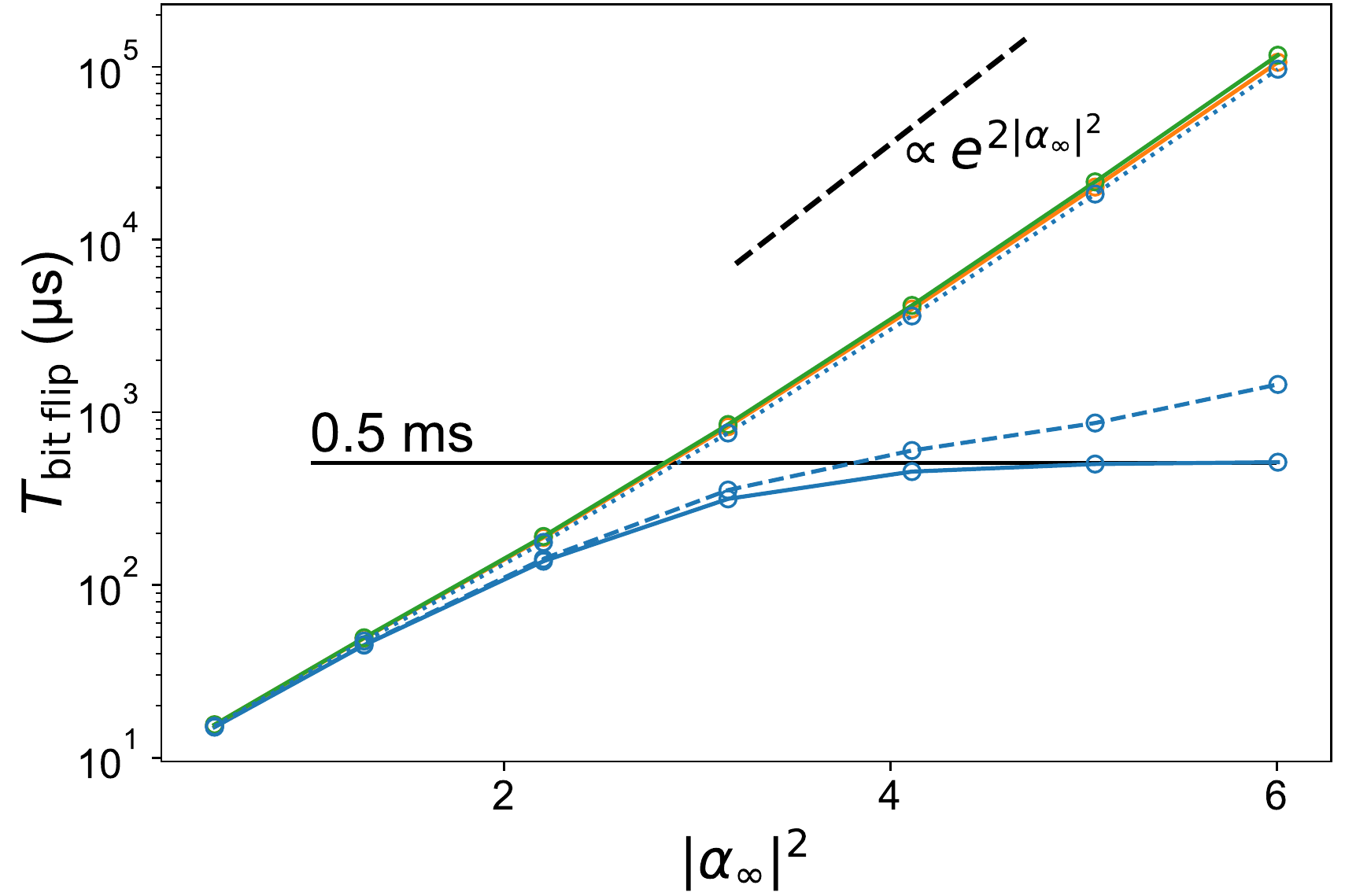}
\caption{Simulated bit-flip time as a function of the cat-qubit size $|\ainf|^2$. We simulate the experiment displayed in Fig.~\ref{fig4} of the main text given the measured system parameters (Tables~\ref{table} and \ref{table0}) with a master equation solver (QuTiP) for various values of $\alpha^2$. The cat-qubit size is given by $|\alpha_\infty|^2 = |\!\braket{\ba^2}\!|$ after a time $t\gg\kde^{-1}$. We simulate three models of increasing complexity: a one, two and three modes model. The green curve corresponds to the simulation of the equivalent dynamics of the cat-qubit resonator alone \eqref{eq:me1}. We get an exponential increase of the bit-flip time with a dependence on the cat size close to the theoretical prediction $\exp(2|\ainf|^2)$ \cite{CohenThesis_supp} (dashed black line is a guide for the eye). The orange curve corresponds to the simulation of the dynamics of the buffer and cat-qubit together \eqref{eq:me2}. The good agreement between these two curves indicates that the adiabatic elimination of the buffer is valid (indeed $g_2\ll \kappa_b$). Finally the blue solid curve is the simulation including the transmon with its thermal occupation \eqref{eq:me3}. In this case, the bit-flip time saturates at around 0.5~ms which is compatible to the experimentally measured value (1~ms). This saturation follows a prior exponential increase where the bit-flip time is multiplied by 3.7 for each added photon (experimentally 4.2). We also simulate for $\chi_{qa} = \chi_{qa, \mathrm{exp}}/3$ (dashed blue line) and $\chi_{qa, \mathrm{exp}}/10$ = 72 kHz (dotted blue line) and as expected, the curve approaches the exponential scaling when $\chi_{qa}$ is low enough.
}
\label{fig_simu}
\end{figure}


\subsection{Bit-flip time simulation}

In the previous part, we gained insight on how single photon loss and detuning affect the cat-qubit protection. In the following, we perform a full master equation simulation of the system with the measured system parameters. The system consists of three relevant modes: the buffer and cat-qubit resonator, and the transmon qubit. We can write the Hamiltonian and loss operators (in the rotating frame for each mode)
\begin{eqnarray}
    \bm{H_3}/\hbar &=& \left(g_2^*(\ba^2-\alpha^2)\bm{b}^\dag+ \mathrm{h.c}\right) - \frac{\chi_{aa}}{2}\ba^{\dag2}\ba^2 - \chi_{qa}\ba^\dag\ba\bm{q}^\dag\bm{q} \label{eq:me3} \\
    \bm{L}_a=\sqrt{\kappa_a}\ba,\,
    \bm{L}_b&=&\sqrt{\kappa_b}\bb ,\,
    \bm{L}_q=\sqrt{\kappa_q(1+n_\mathrm{th})}\bq ,\,
    \bm{L}_{q^\dag}=\sqrt{\kappa_q n_\mathrm{th}}\bq^\dag \nonumber
\end{eqnarray}
where we have (from left to right) in the Hamiltonian, the two-to-one photon exchange factored with a drive on the buffer with strength $\epsilon_d=-g_2^*\alpha^2$, the Kerr of the cat-qubit, the cross-Kerr between the cat-qubit and the transmon. The two last loss operators model the decay rate of the transmon (rate $\kappa_q$), and its thermal occupation measured to be $n_{th}\sim1$\% in the presence of the pump and drive. 
In order to determine the effect of the transmon on the cat-qubit, it is useful to remove the transmon from the simulation. We simulate the dynamics generated by
\begin{eqnarray}
    \bm{H_2}/\hbar &=& \left(g_2^*(\ba^2-\alpha^2)\bm{b}^\dag+ \mathrm{h.c}\right) -\frac{\chi_{aa}}{2}\ba^{\dag2}\ba^2 \label{eq:me2}\\
    \bm{L}_a&=&\sqrt{\kappa_a}\ba,\,
    \bm{L}_b=\sqrt{\kappa_b}\bb \nonumber
\end{eqnarray}
Finally by adiabatically eliminating the buffer \cite{Leghtas2015_supp} we reduce to the following equivalent Hamiltonian and loss operator (provided $g_2\ll\kappa_b$)
\begin{eqnarray}
    \bm{H_1}/\hbar &=& - \frac{\chi_{aa}}{2}\ba^{\dag2}\ba^2 \label{eq:me1}\\
    \bm{L}_a&=&\sqrt{\kappa_a}\ba,\,
    \bm{L}_2=\sqrt{\kde}(\ba^2-\alpha^2) \nonumber
\end{eqnarray}
with $\kappa_2={4|g_2|^2}/{\kappa_b}$.

For each of these models, we numerically solve the master equation for the cat-qubit resonator prepared in state $\ket{+\alpha}$ for various $\alpha$. By fitting the decay of $\braket{\ba}$ to an exponential decay, we extract $T_\mathrm{bit-flip}$ that we reported in Fig.~\ref{fig_simu} (full lines). For the last two models, we recover the exponential increase of the bit-flip time which scales as $\sim\exp(2\alpha^2)$. The three modes model reproduces the saturation we have in the experiment and associates it with the thermal excitation of the transmon. Indeed one transmon excitation detunes the cat-qubit by $\chi$ which exceeds $\kappac/2$ when $\alpha^2<29$, well above cat sizes we could achieve experimentally. In future experiment, we plan on reducing $\chi$. By dividing $\chi$ by 10, we expect to fully circumvent this saturation (dotted lines in Fig.~\ref{fig_simu}).

\begin{figure}[!ht]
\includegraphics[width={510pt}]{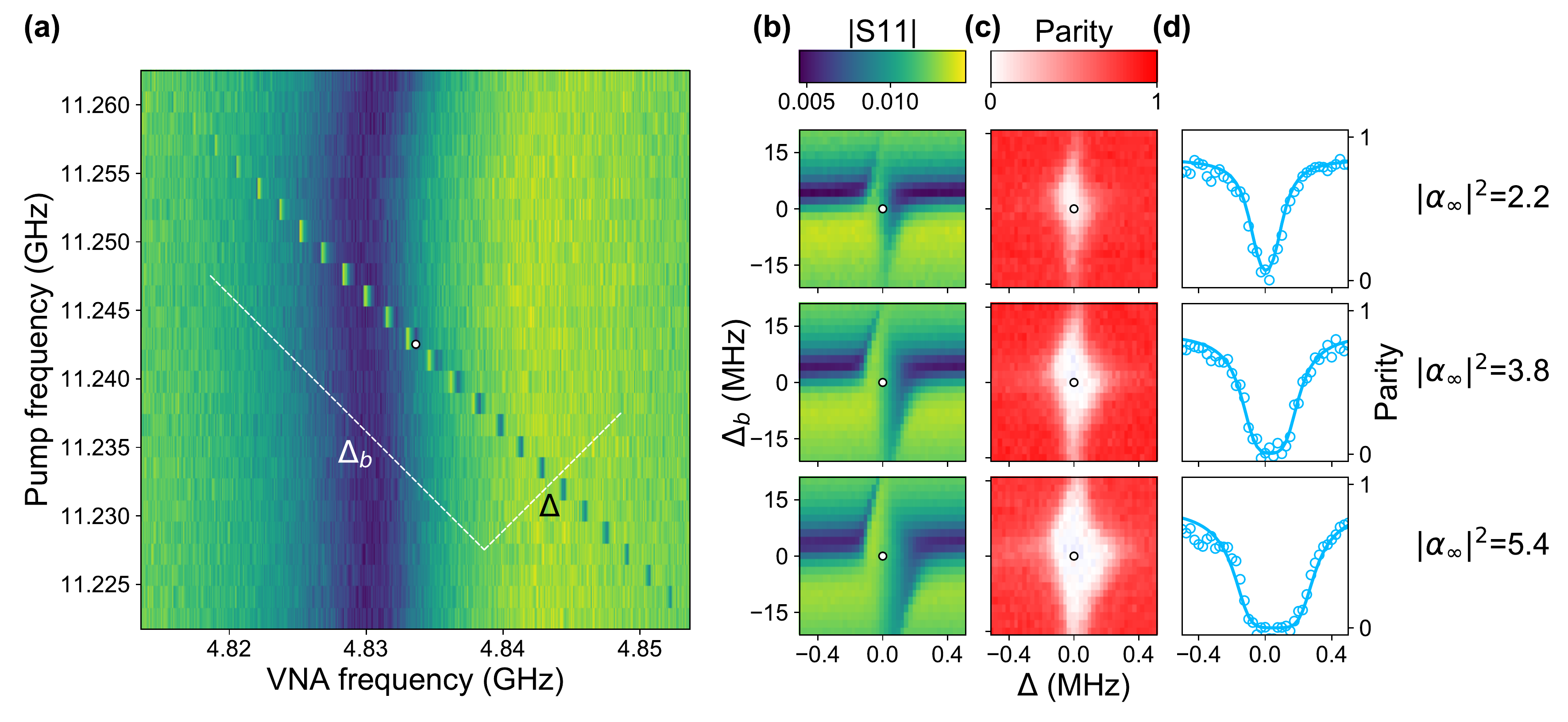}
\caption{\textbf{Tuning the pump and drive frequencies}.
(\textbf{a}) Reflected relative drive amplitude (VNA measurement) as a function of drive frequency (x-axis) and pump frequency (y-axis). When $\omega_p=2\omega_a-\omega_d$, a sharp feature indicates that the two-to-one photon exchange is resonant and as expected, it has a slope $-1$. To observe this feature, we switch to the basis $\Delta = (\Delta_\mathrm{pump}+\Delta_\mathrm{drive})/2$, $\Delta_b=(\Delta_\mathrm{pump}-\Delta_\mathrm{drive})/2$. (\textbf{b,c}) Reflected relative drive amplitude (color) and parity of the cat-qubit resonator (red) as a function of $\Delta$ (x-axis) and $\Delta_b$ (y-axis) for increasing drive amplitude (top to bottom). The drive amplitude is expressed in units of the cat-size $|\ainf|^2$ which is calibrated using the data of Fig~\ref{fig_drive_amp}. (\textbf{c}) When the two-to-one photon exchange is resonant, the cat-qubit resonator is displaced and the parity drops to 0 if we measure after a time greater than $\kun^{-1}$. We  also perform the cat-qubit resonator tomography and verify that the resonator is in a balanced mixture of $\kzero$ and $\kone$. In all these plots, the white circles correspond to the chosen pump and drive frequencies. We verify that for all used drive amplitudes, this point remains centered in the resonant range. Therefore, we do not need to adapt the drive and pump frequencies when increasing the cat size. (\textbf{d}) Cut of the color plot (c) at $\Delta_b=0$ representing the parity (open circle) of the cat-qubit steady state as a function of $\Delta$. The relation \eqref{eq:conf_det} shows that the frequency window over which a non-trivial state is stabilized in the cavity scales as $2\kde|\alpha|^2$. This enables us to determine $\kde$ assuming photon loss is the main loss mechanism. We fit (solid line) the measured parity with the expected steady-state parity (QuTiP) where the two fitting parameters are the parity contrast and $\kde$. We find $\kde/2\pi=40$~kHz.
}
\label{fig_tuning}
\end{figure}

\subsection{Tuning the cat-qubit}
As explained in the main text, the flux point at which the ATS should operate is a saddle point of the buffer frequency map. It is very simple to find experimentally as we do not need to know the full mapping between $(I_{_\Sigma}$, $I_{_\Delta})$ and $(\varphi_{_\Sigma}$, $\varphi_{_\Delta})$ to recognize a saddle point. There are actually two types of saddle points as one can see on Fig.~\ref{fig_flux}, the ones that are tilted to the left and the ones tilted to the right. If the two junctions forming the SQUID of the ATS were perfectly symmetric, these points would be equivalent. Otherwise, the buffer acquires a Kerr non-linearity and the two-points differ by the sign of this Kerr. 

Once we find the buffer and cat-qubit frequencies  we perform two-tone spectroscopy on the buffer (Fig.~\ref{fig_tuning}). A weak tone, referred to as the drive, probes the buffer resonance and the pump is swept in the relevant frequency range (around $2\omega_a-\omega_b$). When the two-to-one exchange occurs between the buffer and the cat-qubit, we observe a sharp feature within the buffer resonance (Fig.~\ref{fig_tuning}a,b). The width of this feature depends on the weak tone strength and more importantly on the pump power. The pump power is pushed until before this feature becomes ill defined, when other non-linear dynamics start to play a significant role. On the cat-qubit side, within this feature the drive combined with the pump should populate the cat-qubit resonator. We check so by measuring the parity of the cat-qubit resonator and verify that it is indeed displaced. We tune the pump and buffer frequency in the middle of the displacement area (Fig.~\ref{fig_tuning}c). The width along $\Delta$ of this region enables us to determine $\kde$.

\begin{figure}[!ht]
\includegraphics[width={510pt}]{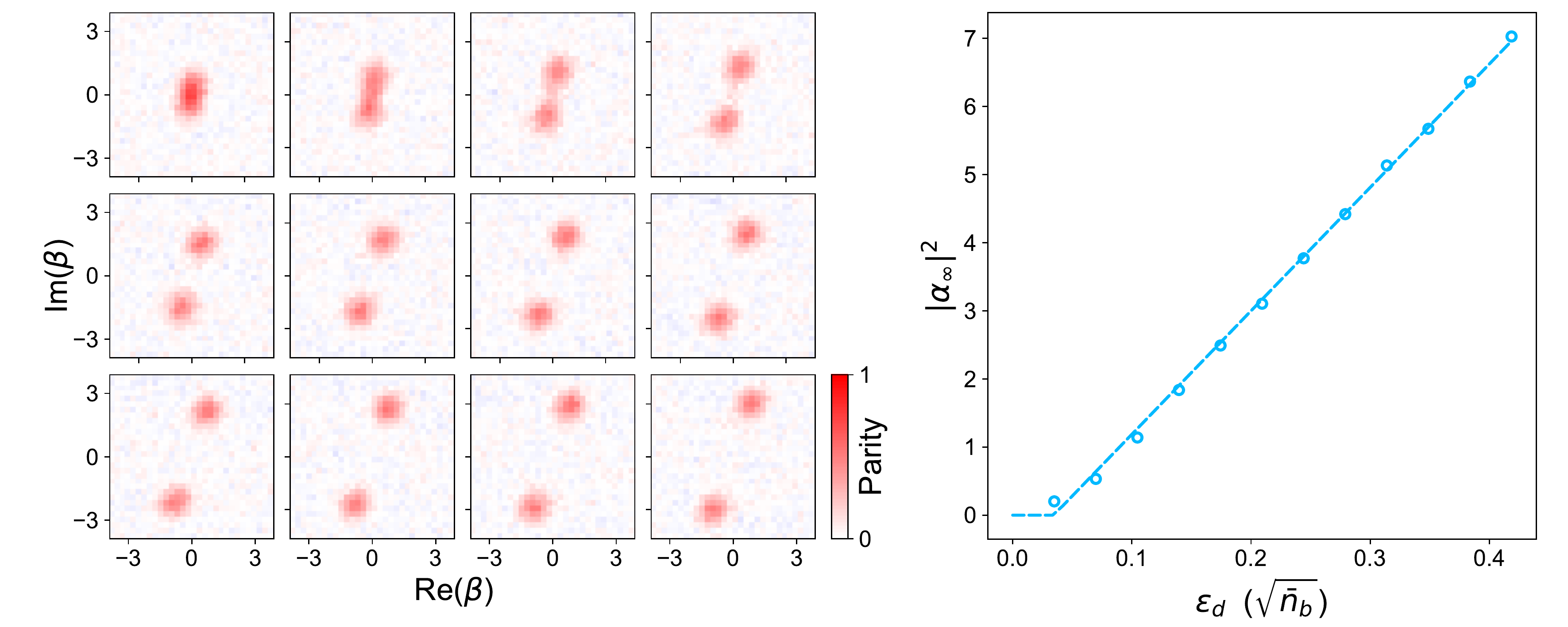}
\caption{\textbf{Increasing the cat-qubit size.} (\textbf{a}) Measured Wigner distribution of the cat-qubit state as a function of drive amplitude (left to right, top to bottom) after a pump and drive pulse duration of $20~\mu$s. (\textbf{b}) Fitted cat size $|\ainf|^2$ (open circles) as a function of the drive amplitude $\epsilon_d$. The drive amplitude is expressed in terms of the square root of the photon number the buffer would contain without the conversion process. For each Wigner distribution of panel (a), we fit a sum of two 2D-Gaussian functions (coherent states) diametrically opposed which are separated by a distance $2|\ainf|$. Note that for simplicity, in the main text, we use $|\alpha|^2$ instead of $|\ainf|^2$. In the presence of single photon loss at rate $\kun$, we expect $|\ainf|^2$ to follow the relation \eqref{eq:ainf} (dashed line): a linear dependence on $\epsilon_d$ shifted by $\kun/(2\kde)$. By fitting this relation to the data, we calibrate the x-axis scaling.}
\label{fig_drive_amp}
\end{figure}

We perform the cat-qubit resonator full tomography after a long ($20~\mathrm{\mu s}\gg\kappa_a^{-1}$) pump and drive pulse (Fig.~\ref{fig_drive_amp}a) and we set the drive amplitude to produce the desired cat size (Fig.~\ref{fig_drive_amp}b). The experiment is now tuned and the cat-qubit characteristics ($T_\text{bit-flip}, \Gamma_\text{phase-flip}$, time evolution of the Wigner function) can be measured.

\bibliographystyle{ieeetr}

\begin{thebibliography}{40}

\bibitem{Shor1995}
P.~Shor, ``Scheme for reducing decoherence in quantum memory,'' {\em Phys. Rev.
  A}, vol.~52, pp.~2493--2496, 1995.

\bibitem{Steane1996}
A.~Steane, ``Error correcting codes in quantum theory,'' {\em Phys. Rev. Lett},
  vol.~77, no.~5, 1996.

\bibitem{Fowler2012}
A.~G. Fowler, M.~Mariantoni, J.~M. Martinis, and A.~N. Cleland, ``Surface
  codes: Towards practical large-scale quantum computation,'' {\em Phys. Rev.
  A.}, vol.~86, p.~032324, Sept. 2012.

\bibitem{Fluhmann2019}
C.~Fl\"{u}hmann, T.~L. Nguyen, M.~Marinelli, V.~Negnevitsky, K.~Mehta, and
  J.~P. Home, ``Encoding a qubit in a trapped-ion mechanical oscillator,'' {\em
  Nature}, vol.~566, pp.~513--517, Feb. 2019.

\bibitem{Campagne2019}
P.~Campagne-Ibarcq~\emph{et al.} 2019.
\newblock In preparation.

\bibitem{Brooks2013}
P.~Brooks, A.~Kitaev, and J.~Preskill, ``Protected gates for superconducting
  qubits,'' {\em Physical Review A}, vol.~87, May 2013.

\bibitem{Albrecht2016}
S.~M. Albrecht, A.~P. Higginbotham, M.~Madsen, F.~Kuemmeth, T.~S. Jespersen,
  J.~Nyg{\aa}rd, P.~Krogstrup, and C.~M. Marcus, ``Exponential protection of
  zero modes in majorana islands,'' {\em Nature}, vol.~531, pp.~206--209, Mar.
  2016.

\bibitem{Lin2018}
Y.-H. Lin, L.~B. Nguyen, N.~Grabon, J.~San~Miguel, N.~Pankratova, and V.~E.
  Manucharyan, ``Demonstration of protection of a superconducting qubit from
  energy decay,'' {\em Phys. Rev. Lett.}, vol.~120, p.~150503, Apr 2018.

\bibitem{Earnest2018}
N.~Earnest, S.~Chakram, Y.~Lu, N.~Irons, R.~K. Naik, N.~Leung, L.~Ocola, D.~A.
  Czaplewski, B.~Baker, J.~Lawrence, J.~Koch, and D.~I. Schuster, ``Realization
  of a $\mathrm{\ensuremath{\Lambda}}$ system with metastable states of a
  capacitively shunted fluxonium,'' {\em Phys. Rev. Lett.}, vol.~120,
  p.~150504, Apr 2018.

\bibitem{Smith2019}
W.~C. Smith, A.~Kou, X.~Xiao, U.~Vool, and M.~H. Devoret, ``Superconducting
  circuit protected by two-cooper-pair tunneling,'' {\em arXiv:1905.01206},
  2019.

\bibitem{Puri2017}
S.~Puri, S.~Boutin, and A.~Blais, ``Engineering the quantum states of light in
  a kerr-nonlinear resonator by two-photon driving,'' {\em npj Quantum
  Information}, vol.~3, no.~1, p.~18, 2017.

\bibitem{Wolinsky1988}
M.~Wolinsky and H.~J. Carmichael, ``Quantum noise in the parametric oscillator:
  From squeezed states to coherent-state superpositions,'' {\em Phys. Rev.
  Lett.}, vol.~60, pp.~1836--1839, May 1988.

\bibitem{Leghtas2015}
Z.~Leghtas, S.~Touzard, I.~M. Pop, A.~Kou, B.~Vlastakis, A.~Petrenko, K.~M.
  Sliwa, A.~Narla, S.~Shankar, M.~J. Hatridge, M.~Reagor, L.~Frunzio, R.~J.
  Schoelkopf, M.~Mirrahimi, and M.~H. Devoret, ``Confining the state of light
  to a quantum manifold by engineered two-photon loss,'' {\em Science},
  vol.~347, no.~6224, pp.~853--857, 2015.

\bibitem{Touzard2018a}
S.~Touzard, A.~Grimm, Z.~Leghtas, S.~O. Mundhada, P.~Reinhold, C.~Axline,
  M.~Reagor, K.~Chou, J.~Blumoff, K.~M. Sliwa, S.~Shankar, L.~Frunzio, R.~J.
  Schoelkopf, M.~Mirrahimi, and M.~H. Devoret, ``Coherent oscillations inside a
  quantum manifold stabilized by dissipation,'' {\em Phys. Rev. X}, vol.~8,
  p.~021005, Apr 2018.

\bibitem{Guillaud2019}
J.~Guillaud and M.~Mirrahimi, ``Repetition cat-qubits: fault-tolerant quantum
  computation with highly reduced overhead,'' {\em arXiv:1904.09474}, 2019.

\bibitem{Mirrahimi2014}
M.~Mirrahimi, Z.~Leghtas, V.~V. Albert, S.~Touzard, R.~J. Schoelkopf, L.~Jiang,
  and M.~H. Devoret, ``Dynamically protected cat-qubits: a new paradigm for
  universal quantum computation,'' {\em New J. Phys.}, vol.~16, no.~4,
  p.~045014, 2014.

\bibitem{Grimm2019}
A.~Grimm~\emph{et al.} 2019.
\newblock In preparation.

\bibitem{Puri2019}
S.~Puri, L.~St-Jean, J.~A. Gross, A.~Grimm, N.~E. Frattini, P.~S. Iyer,
  A.~Krishna, S.~Touzard, L.~Jiang, A.~Blais, S.~T. Flammia, and S.~M. Girvin,
  ``Bias-preserving gates with stabilized cat qubits,'' {\em arXiv:1905.00450},
  2019.

\bibitem{Leghtas2013}
Z.~Leghtas, G.~Kirchmair, B.~Vlastakis, R.~J. Schoelkopf, M.~H. Devoret, and
  M.~Mirrahimi, ``{Hardware-Efficient Autonomous Quantum Memory Protection},''
  {\em Physical Review Letters}, vol.~111, p.~120501, sep 2013.

\bibitem{supplement}
 {\em Supplementary Materials}.

\bibitem{Gottesman2001}
D.~Gottesman, A.~Kitaev, and J.~Preskill, ``Encoding a qubit in an
  oscillator,'' {\em Phys. Rev. A}, vol.~64, p.~012310, 2001.

\bibitem{Carmichael2007}
H.~J. Carmichael, {\em Statistical Methods in Quantum Optics 2}.
\newblock {Springer}, 2007.

\bibitem{Sank2016}
D.~Sank, Z.~Chen, M.~Khezri, J.~Kelly, R.~Barends, B.~Campbell, Y.~Chen,
  B.~Chiaro, A.~Dunsworth, A.~Fowler, E.~Jeffrey, E.~Lucero, A.~Megrant,
  J.~Mutus, M.~Neeley, C.~Neill, P.~J.~J. O'Malley, C.~Quintana, P.~Roushan,
  A.~Vainsencher, T.~White, J.~Wenner, A.~N. Korotkov, and J.~M. Martinis,
  ``Measurement-induced state transitions in a superconducting qubit: Beyond
  the rotating wave approximation,'' {\em Phys. Rev. Lett.}, vol.~117,
  p.~190503, Nov 2016.

\bibitem{Gao2018}
Y.~Y. Gao, B.~J. Lester, Y.~Zhang, C.~Wang, S.~Rosenblum, L.~Frunzio, L.~Jiang,
  S.~M. Girvin, and R.~J. Schoelkopf, ``Programmable interference between two
  microwave quantum memories,'' {\em Phys. Rev. X}, vol.~8, p.~021073, Jun
  2018.

\bibitem{LescannePRApp2019}
R.~Lescanne, L.~Verney, Q.~Ficheux, M.~H. Devoret, B.~Huard, M.~Mirrahimi, and
  Z.~Leghtas, ``Escape of a driven quantum josephson circuit into unconfined
  states,'' {\em Phys. Rev. Applied}, vol.~11, p.~014030, Jan 2019.

\bibitem{Wang2014}
C.~Wang, Y.~Y. Gao, I.~M. Pop, U.~Vool, C.~Axline, T.~Brecht, R.~W. Heeres,
  L.~Frunzio, M.~H. Devoret, G.~Catelani, L.~I. Glazman, and R.~J. Schoelkopf,
  ``Measurement and control of quasiparticle dynamics in a superconducting
  qubit,'' {\em Nature Communications}, vol.~5, pp.~5836 EP --, 12 2014.

\bibitem{Vrajitoarea2018}
A.~Vrajitoarea, Z.~Huang, P.~Groszkowski, J.~Koch, and A.~A. Houck, ``Quantum
  control of an oscillator using stimulated nonlinearity,'' 2018.

\bibitem{VerneyPRApp2019}
L.~Verney, R.~Lescanne, M.~H. Devoret, Z.~Leghtas, and M.~Mirrahimi,
  ``Structural instability of driven josephson circuits prevented by an
  inductive shunt,'' {\em Phys. Rev. Applied}, vol.~11, p.~024003, Feb 2019.

\bibitem{Touzard2018b}
S.~Touzard, A.~Kou, N.~E. Frattini, V.~V. Sivak, S.~Puri, A.~Grimm, L.~Frunzio,
  S.~Shankar, and M.~H. Devoret, ``Gated conditional displacement readout of
  superconducting qubits,'' {\em Phys. Rev. Lett.}, vol.~122, p.~080502, Feb
  2019.

\bibitem{Raimond2006}
J.~Raimond and S.~Haroche, ``{Exploring the Quantum},'' {\em Oxford University
  Press}, vol.~82, no.~1, p.~86, 2006.

\bibitem{Reagor2016}
M.~Reagor, W.~Pfaff, C.~Axline, R.~W. Heeres, N.~Ofek, K.~Sliwa, E.~Holland,
  C.~Wang, J.~Blumoff, K.~Chou, M.~J. Hatridge, L.~Frunzio, M.~H. Devoret,
  L.~Jiang, and R.~J. Schoelkopf, ``Quantum memory with millisecond coherence
  in circuit qed,'' {\em Phys. Rev. B}, vol.~94, p.~014506, Jul 2016.

\end{thebibliography}

\begin{thebibliography}{40}
\bibitem{Leghtas2015_supp}
Z.~Leghtas, S.~Touzard, I.~M. Pop, A.~Kou, B.~Vlastakis, A.~Petrenko, K.~M.
  Sliwa, A.~Narla, S.~Shankar, M.~J. Hatridge, M.~Reagor, L.~Frunzio, R.~J.
  Schoelkopf, M.~Mirrahimi, and M.~H. Devoret, ``Confining the state of light
  to a quantum manifold by engineered two-photon loss,'' {\em Science},
  vol.~347, no.~6224, pp.~853--857, 2015.

\bibitem{Touzard2018a_supp}
S.~Touzard, A.~Grimm, Z.~Leghtas, S.~O. Mundhada, P.~Reinhold, C.~Axline,
  M.~Reagor, K.~Chou, J.~Blumoff, K.~M. Sliwa, S.~Shankar, L.~Frunzio, R.~J.
  Schoelkopf, M.~Mirrahimi, and M.~H. Devoret, ``Coherent oscillations inside a
  quantum manifold stabilized by dissipation,'' {\em Phys. Rev. X}, vol.~8,
  p.~021005, Apr 2018.

\bibitem{Guillaud2019_supp}
J.~Guillaud and M.~Mirrahimi, ``Repetition cat-qubits: fault-tolerant quantum
  computation with highly reduced overhead,'' {\em arXiv:1904.09474}, 2019.

\bibitem{Mirrahimi2014_supp}
M.~Mirrahimi, Z.~Leghtas, V.~V. Albert, S.~Touzard, R.~J. Schoelkopf, L.~Jiang,
  and M.~H. Devoret, ``Dynamically protected cat-qubits: a new paradigm for
  universal quantum computation,'' {\em New J. Phys.}, vol.~16, no.~4,
  p.~045014, 2014.
  
\bibitem{Frattini2018_supp}
N.~E. Frattini, V.~V. Sivak, A.~Lingenfelter, S.~Shankar, and M.~H. Devoret,
  ``Optimizing the nonlinearity and dissipation of a snail parametric amplifier
  for dynamic range,'' {\em Phys. Rev. Applied}, vol.~10, p.~054020, Nov 2018.
  
\bibitem{LescannePRApp2019_supp}
R.~Lescanne, L.~Verney, Q.~Ficheux, M.~H. Devoret, B.~Huard, M.~Mirrahimi, and
  Z.~Leghtas, ``Escape of a driven quantum josephson circuit into unconfined
  states,'' {\em Phys. Rev. Applied}, vol.~11, p.~014030, Jan 2019.

\bibitem{VerneyPRApp2019_supp}
L.~Verney, R.~Lescanne, M.~H. Devoret, Z.~Leghtas, and M.~Mirrahimi,
  ``Structural instability of driven josephson circuits prevented by an
  inductive shunt,'' {\em Phys. Rev. Applied}, vol.~11, p.~024003, Feb 2019.


\bibitem{Reed2010_supp}
M.~D. Reed, B.~R. Johnson, A.~A. Houck, L.~DiCarlo, J.~M. Chow, D.~I. Schuster,
  L.~Frunzio, and R.~J. Schoelkopf, ``Fast reset and suppressing spontaneous
  emission of a superconducting qubit,'' {\em Appl. Phys. Lett.}, vol.~96,
  p.~203110, May 2010.

\bibitem{CohenThesis_supp}
J.~Cohen, {\em {Autonomous quantum error correction with superconducting
  qubits}}.
\newblock Theses, {PSL Research University}, Feb. 2017.

\bibitem{Mirrahimi2019_supp}
M.~Mirrahimi, ``Quantum error correction with cat qubits,'' {\em Les Houches
  Session, Quantum Information Machines}, 2019.

\end{thebibliography}

\end{document}